\documentclass[11pt,letterpaper,twocolumn]{article} 

\usepackage[utf8]{inputenc}
\usepackage[english]{babel}
\usepackage{float}
\usepackage{xcolor}
\usepackage{verbatim}
\usepackage{mwe}
\usepackage{charter}
\usepackage{afterpage}
\usepackage{amsmath}
\usepackage{amsfonts} %
\usepackage{amssymb}
\usepackage{dsfont}
\usepackage{mathrsfs}
\usepackage{appendix}
\usepackage{ragged2e}
\usepackage{array}
\usepackage{braket}
\usepackage{etoolbox}
\usepackage{fancyhdr}
\usepackage{booktabs}
\usepackage{arydshln}
\usepackage[nohug,heads=LaTeX]{diagrams}  
\diagramstyle[labelstyle=\scriptstyle]
\usepackage[justification=justified,singlelinecheck=false,labelfont=bf,format=plain]{caption}
\usepackage[justification=justified,singlelinecheck=false,labelfont=bf,format=plain]{subcaption}
\usepackage{enumitem}
\usepackage[bottom=2.5cm,top=2.0cm,left=2.0cm,right=2.0cm]{geometry}
\usepackage{graphicx}
\usepackage{indentfirst}
\usepackage{mathtools}
\usepackage{multirow}
\usepackage{pdfpages}
\usepackage{tikz}
\usepackage{lmodern} 

\usepackage{subfiles}
\usepackage[compact]{titlesec}
\usepackage{blindtext}
\usepackage{stfloats}
\usepackage{lipsum} 

\usepackage{hyperref}
\usepackage{cleveref}

\usepackage[square,numbers]{natbib}

\usepackage{csquotes}

\newtheorem{theorem}{Theorem}

\newtheorem{algorithm}[theorem]{Algorithm}

\newtheorem{corollary}[theorem]{Corollary}

\newtheorem{definition}[theorem]{Definition}
\newtheorem{example}[theorem]{Example}

\newtheorem{lemma}[theorem]{Lemma}

\newtheorem{remark}[theorem]{Remark}

\newtheorem{observation}[theorem]{Observation/Theorem}
\newenvironment{proof}[1][Proof]{\textbf{#1.} }{\ \rule{0.5em}{0.5em}}

\DeclareMathOperator{\Tr}{Tr}

\DeclareMathOperator{\diag}{Diag}
\DeclareMathOperator{\ind}{ind}

\newcolumntype{L}[1]{>{\raggedright\let\newline\\\arraybackslash\hspace{0pt}}m{#1}}
\newcolumntype{C}[1]{>{\centering\let\newline\\\arraybackslash\hspace{0pt}}m{#1}}
\newcolumntype{R}[1]{>{\raggedleft\let\newline\\\arraybackslash\hspace{0pt}}m{#1}}

    \setlist[itemize,1]{label=$\bullet$}
    \setlist[itemize,2]{label=$\circ$}
    \setlist[itemize,3]{label=$-$}
    \setlist{nosep}

\titlelabel{\thetitle.\quad}

\makeatletter
\patchcmd{\headrule}{\hrule}{\color{black}\hrule}{}{} 
\patchcmd{\footrule}{\hrule}{\color{black}\hrule}{}{} 
\makeatother

\definecolor{blueM}{cmyk}{1.0,0.49,0.0,0.47}

\begin{document}

\hspace{30pt}
\begin{minipage}{0.75\textwidth}
\vspace{4mm}
		\begin{center}
		\Large{\textbf{Low-dimensional polaritonics:\\ Emergent non-trivial topology on exciton-polariton simulators}}
		\end{center}
    \large{Author: Konstantin Rips$^{1}$} 
		\vspace{3mm} \newline
    $^1$\fontsize{0.35cm}{0.5cm}\selectfont \textit{Physics Department, Faculty of Science and Technology, Lancaster University,
    Bailrigg, Lancaster LA1 4YW, United Kingdom} 
    \vspace{1mm} 
		
\end{minipage}

\small
\vspace{11pt}

\begin{center} 
    \begin{minipage}{0.9\textwidth}
        \noindent \textbf{Abstract:} Polaritonic lattice configurations in dimensions D=2 are used as simulators of topological phases, based on symmetry class A Hamiltonians. Numerical and topological studies are performed in order to characterise the bulk topology of insulating phases, which is predicted to be connected to non-trivial edge mode states on the boundary. By using spectral–flattened Hamiltonians on specific lattice geometries with time–reversal symmetry breaking, e.g. Kagome lattice, I obtain maps from the Brillouin zone into Grassmannian spaces, which are further investigated by the topological method of space fibrations. Numerical evidence reveals a connection between the sum of valence band Chern numbers and the index of the projection operator onto the valence band states. Along these lines, I discover an index formula which resembles other index theorems and the classical result of Atiyah-Singer, but without any Dirac operator and from a different perspective. Through a combination of different tools, in particular homotopy and homology-cohomology duality, we provide a comprehensive mathematical framework, which fully addresses the source and structure of topological phases in coupled polaritonic array systems. 
Based on these results, it becomes possible to infer further designs and models of two-dimensional single sheet Chern insulators, implemented as polaritonic simulators.

        \vspace{4mm}
    
    \end{minipage}
    
\end{center}



\onecolumn   

\section{Introduction}
\justify
Two-dimensional exciton-polariton platforms have received quite some attraction by the condensed matter science community due to the possibility of engineering hybrid light-matter crystals, e.g. topological insulators. A characteristic feature of these materials is a topological gapped bulk spectrum, which is connected to edge mode states on the boundary through the so called bulk-boundary correspondence. Due to their topological origin, these states are robust against perturbations, show unidirectional transport, and are insensitive to backscattering into the bulk. In contrast to their electronic counterparts, which exhibit mostly $\mathds Z_2$ type of phases, the bosonic phases are characterized by integer topological numbers. The bosons considered here are known as polaritons, or exciton-polaritons, which are quasi-particles emerging from photons and quantum well excitons entering the strong coupling regime in engineered semiconductor micro-cavities. As bosonic particles, polaritons can undergo a phase transition into a Bose-Einstein condensate (BEC). Furthermore, they exhibit various other collective quantum phenomena, such as lasing and superfluidity \cite{Suchomel, Amo}. There exist some common aspects to atomic BECs - however, atomic BECs are strongly related to thermodynamic equilibrium, whereas polaritonic configurations are characterised by a non-equilibrium setting due to their decay into photons and thus, short lifetime. For a stable population configuration, the system must be frequently restocked from a pump source.
Our actual interest, however, is in nano-fabricated polaritonic metamaterials which mimic topological insulators; these are materials which behave as insulators inside the bulk, but develop edge states on their surface as a result of the bulk-boundary correspondence. The topologically protected edge modes are insensitive to local perturbations or defects because of the bulk topology. Long range spatial coherence \cite{Deng} makes polariton condensates an attractive candidate for engineering landscapes of topologically non-trivial phases compared to purely electronic systems. Several techniques for creating trapping potentials for polariton condensates, similar to optical lattices for atoms, have been proposed. This paves the way to novel applications in quantum simulation, optimization and the design of quantum devices \cite{Berloff, Cohen}.

\justify
\par \vspace{5mm}   



\section{Preliminaries and theoretical extensions on lattice systems}

This section introduces some formalism of importance to quantum lattice models in condensed matter physics. Moreover, we present some underlying topological aspects from a novel point of view. The homological-cohomological relation between lattice structure and coset space of the Hamiltonian is considered, and implications thereof are discussed.

\subsection{Structure of Lattice-Hamiltonian}

The general lattice Hamiltonian we are studying is given by  
\begin{align}
\hat H&=\sum_{i} U_i\hat a_i^{\dagger}\hat a_i + \sum_{\langle i,j\rangle} \kappa_{ij}\hat a^{\dagger}_i\hat a_j + \sum_{\langle\langle i,j\rangle\rangle}\kappa_{ij}\hat a^{\dagger}_i\hat a_j + \cdots ,\label{GHam_Latt} \\
a^{\dagger}_i\ket{0}&=\ket{\textsf{\tiny{particle at site i}}}, 
\end{align}
in the second quantization formalism. The sum runs over \emph{nearest-neighbour} (\emph{nn}) and \textit{next-nearest-neighbour pairs} (\textit{nnn}), denoted by $\langle i,j\rangle$ and $\langle\langle i,j\rangle\rangle$, respectively. $\hat a^{\dagger}_i$ and $\hat a_j$ represent the creation and annihilation operators of \textsl{particles} at sites $i$ and $j$, respectively. The particles can be atoms, electrons, or quasi-particles, e.g. polaritons. However, care must be taken as to whether they obey Fermi-Dirac or Bose-Einstein statistics. $\ket{0}$ is a generic vacuum state. Note that we suppress other properties of the particles, such as polarization, spin etc., which could be actually present. We refer to the matrix elements $\kappa_{ij}$ as hopping amplitudes, which can be written as space integrals of overlapping orbitals for two neighbouring sites $i$, $j$,
\begin{equation}
\kappa_{ij}=\bra{i}\hat{\mathcal J}\ket{j}= \int\int\phi^*_i(\textbf{x})\mathcal J(\textbf{x},\textbf{y})\phi_j(\textbf{y})\,d\textbf{x}\,d\textbf{y},
\end{equation}
$\hat{\mathcal J}$ denotes the operator inducing the particle hopping process. In the presence of an external potential, the above Hamiltonian also includes on-site terms $U_i\hat a_i^{\dagger}\hat a_i$. Moreover, it is possible to include two-body-interactions in the form
\begin{equation}
\hat V=\frac{1}{2}\sum_{i,j,k,m}V_{ijkm}\hat a_{i}^{\dagger}\hat a_{j}^{\dagger}\hat a_{m}\hat a_{k},
\end{equation}
which is added to the non-interacting lattice Hamiltonian. In particular, one can show that polariton graphs possess on-site interactions, which also arise in other quantum lattice models, e.g. the bosonic or fermionic Hubbard model. A plethora of lattice Hamiltonians has recently been designed for both Bose and Fermi gases \cite{Lewenstein}, some of which even reveal the phenomenon of fermion fractionalization \cite{Ruostek}. 

\subsection{Prominent Lattices and Brillouin Zone Topology} 

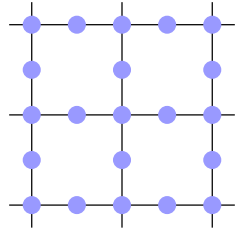
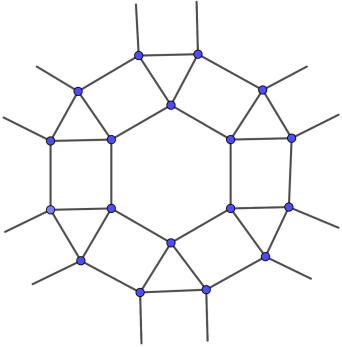
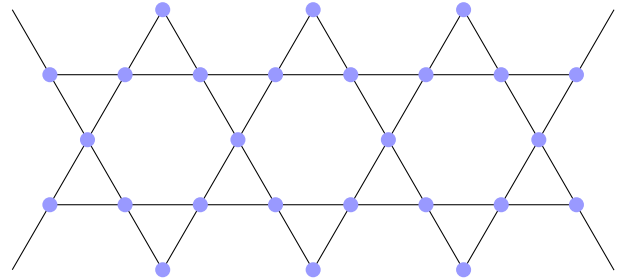
\begin{figure}[H]
\centering
\subfloat[Lieb Lattice]{\scalebox{0.6}
{\begin{tikzpicture}[scale=1] 
\tikzstyle{every node} = [circle]
\draw[thick] (-1,-1.5) -- (-1,3.5);
\draw[thick] (1,-1.5) -- (1,3.5);
\draw[thick] (3,-1.5) -- (3,3.5);
\draw[thick] (-1.5,-1) -- (3.5,-1);
\draw[thick] (-1.5,1) -- (3.5,1);
\draw[thick] (-1.5,3) -- (3.5,3);
\foreach \x in {-1,0,1,2,3}
 \foreach \y in {-1,1,3}
  \fill[color=blue!40](\x,\y) circle (0.2);
\foreach \x in {-1,1,3}
 \foreach \y in {0,2}
  \fill[color=blue!40](\x,\y) circle (0.2);	
\end{tikzpicture}}}

\subfloat[Ruby (Diamond) Lattice]{\scalebox{0.6}
{\includegraphics[width=.5\textwidth]{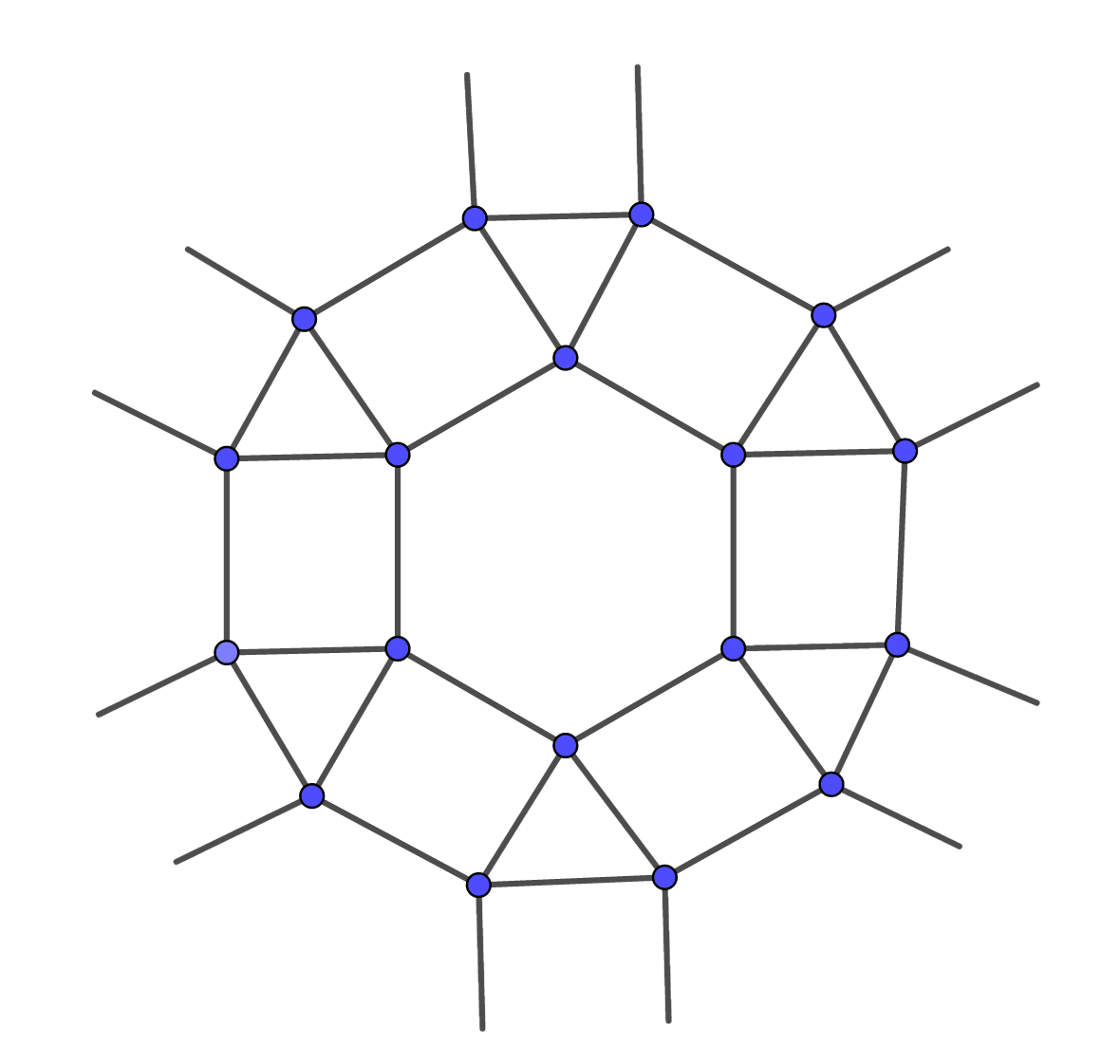}}}
\hfill
\subfloat[Kagome Lattice]{\scalebox{0.5}
{\begin{tikzpicture}
\tikzstyle{every node} = [circle]
\path[draw,thick] (-4,0)--(2,0)--(10,0);
\path[draw,thick] (-4,3.46)--(2,3.46)--(10,3.46);
\path[draw,thick] (-5,5.19)--(-3,1.73)--(-1,-1.74);
\path[draw,thick] (-1,5.19)--(1,1.73)--(3,-1.74);
\path[draw,thick] (3,5.19)--(5,1.73)--(7,-1.74);
\path[draw,thick] (7,5.19)--(9,1.73)--(11,-1.74);
\path[draw,thick] (-5,-1.73)--(-3,1.73)--(-1,5.19);
\path[draw,thick] (-1,-1.73)--(1,1.73)--(3,5.19);
\path[draw,thick] (3,-1.73)--(5,1.73)--(7,5.19);
\path[draw,thick] (7,-1.73)--(9,1.73)--(11,5.19);
\foreach \b in {-2,0,...,5}
  \fill[color=blue!40](2*\b,0) circle (0.2);
\foreach \w in {-1,1,...,5}
  \fill[color=blue!40](2*\w,0) circle (0.2);	
\foreach \w in {-2,0,...,5}
  \fill[color=blue!40](2*\w,3.46) circle (0.2);
\foreach \b in {-1,1,...,5}
  \fill[color=blue!40](2*\b,3.46) circle (0.2);	
\foreach \r in {-1.5,0.5,...,5}
  \fill[color=blue!40](2*\r,1.73) circle (0.2);		
\foreach \ru in {-0.5,1.5,...,5}
  \fill[color=blue!40](2*\ru,-1.73) circle (0.2);			
\foreach \ro in {-0.5,1.5,...,5}
  \fill[color=blue!40](2*\ro,5.19) circle (0.2);	
\end{tikzpicture}}}
\caption{Snapshots of prominent lattice structures, beyond the well known square, triangular or hexagonal geometry, which can exhibit topological properties. Lattice sites are depicted by blue dots and the links between them indicate non--vanishing hopping amplitudes. In all three cases, one can define a non-trivial unit cell which consists of multiple levels. Consequently, this property creates a multiple band structure over the momentum space.}
\label{fig:Lattices}
\end{figure}

Hamiltonians designed on 2D periodic lattices, as in \cref{fig:Lattices}, share a crystallographic feature: There exists a 2D Bravais lattice $\Lambda=\{\textbf{R}(m,n):=m\textbf{a}_1+n\textbf{a}_2|(m,n)\in \mathbb Z^2\}$ such that unit cells can be translated by elements of $\Lambda$ while leaving the lattice geometry invariant. In practical applications, $(m,n)$ denotes the unit cell, and $\textbf{a}_1/\textbf{a}_2$ are the primitive vectors generating the lattice $\Lambda$. These crystallographic concepts can be extended to higher dimensions.

\paragraph{Methods - Fourier Spectroscopy.} 
Let $\hat H$ be a Hamiltonian \cref{GHam_Latt} which is given on a periodic 2D lattice, e.g. as in \cref{fig:Lattices}. It acts on the Hilbert space $\mathfrak H_{ext}\otimes \mathfrak H_{int}$, where $\mathfrak H_{ext}$ refers to the external space of the underlying lattice structure and $\mathfrak H_{int}$ is the space of internal degrees of freedom (spin, polarization), respectively. Translational invariance allows for a Fourier transformation $\ket{\textbf{k}}=\frac{1}{\sqrt{N}}\sum_{(m,n)\in\mathbb Z^2}e^{i\textbf{k}\cdot\textbf{R}(m,n)}\ket{(m,n)}$ with $\ket{(m,n)}\in \mathfrak H_{ext}$. By this procedure, we obtain a map $\textbf{k}\rightarrow \hat H(\textbf{k})=\bra{\textbf{k}}\hat H\ket{\textbf{k}}$ from the momentum space into the parameter space $\mathcal M_H$ of the Hamiltonian. In the single-particle picture, the parameter space $\mathcal M_H$ is determined by the behaviour of the system under time-reversal (TR), particle-hole (PH) and sublattice (SL) symmetry. All possibilities for $\mathcal M_H$ have been classified by Altland and Zirnbauer, as shown in \cref{fig:Altl_Zirn}. Moreover, periodicity implies the existence of a 2D reciprocal lattice $\Lambda^\ast$ on the (quasi-)momentum space such that invariance under mappings $\textbf{k}\rightarrow \textbf{k}+\textbf{Q}$ with $\textbf{Q}\in \Lambda^\ast$ is guaranteed; $\hat H(\textbf{k}+\textbf{Q})=\hat H(\textbf{k})$. We identify a fundamental region of the momentum space, called the Brillouin zone ($\mathcal{BZ}$), from which the whole plane can be reconstructed by action of the lattice $\Lambda^\ast$. On a formal level one defines the action $\Lambda^\ast\times \mathbb R^2\rightarrow \mathbb R^2$, $(\textbf{Q},\textbf{k})\mapsto\textbf{k}+\textbf{Q}$, where $\Lambda^\ast$ operates as the finitely generated abelian group $\mathbb Z^2$ on the momentum space. One can demonstrate that physically relevant topology is encoded in the maps we have just generically described, i.e.
\begin{equation}
\mathcal{BZ} \rightarrow \mathcal M_H, \quad \textbf{k}\mapsto \hat H(\textbf{k}),
\end{equation}
where $\mathcal M_H$ is one of the coset spaces in \cref{fig:Altl_Zirn}.

\paragraph{The Lieb, Kagome and Ruby Lattice.}
We briefly point out the most remarkable features of 2D lattices shown in \cref{fig:Lattices} with respect to itinerant electrons. By mapping the lattice systems onto a tight-binding model, which is basically a Hamiltonian given in \cref{GHam_Latt}, we get the following pictures: 1) The electronic Lieb lattice resembles a band structure with 2 conic bands touching the third flat band situated in the middle of the spectrum \cite{Weeks}. For instance, electrons hopping on sites of a two-dimensional Lieb lattice, and also [three-dimensional edge centered cubic] perovskite lattice are shown to form topologically non-trivial insulating phases when spin-orbit coupling is included. 2) The picture of the electronic Kagome lattice bands is similar to \cref{fig:Kagome1} b), however with striking differences to a polaritonic system \cite{Guo}. 3) The ruby lattice shows a more complex picture of 6 bands in total \cite{Hu}, for which various hopping parameter implementations have been studied.
All three examples share a common property: spin-orbit (SO) induced (nnn)-coupling of the form $it \sum_{\langle\langle i,j\rangle\rangle\alpha\beta }(\textbf{e}_{ij}^1\times \textbf{e}_{ij}^2)\hat a^{\dagger}_{i\alpha}\sigma_{\alpha\beta}\hat a_{j\beta}$ allows for a gap opening mechanism in the spectrum which drives the system into an insulating phase.
Since these spin-orbit couplings preserve time-reversal symmetry $\hat{\mathcal T}$, we must have $\hat{\mathcal T}^2=-\mathds 1$ for fermions. According to the Altland-Zirnbauer classification, this condition puts considerable restriction on available options for coset spaces of the Hamiltonians. Hence, by examining the potential topologies, one expects to find $\mathbb Z_2$-valued insulating phases in $D=2$, as has been indeed confirmed in all the above cases. In contrast, specific arrangements of polaritonic systems provide a platform for breaking time reversal symmetry (TRS) in two dimensions, and thereby engineering Chern insulators. 


\subsubsection{Topological Analysis}

\paragraph{Brillouin zone $\mathcal{BZ}$.} 
The action $\Lambda^\ast\times \mathbb R^2\rightarrow \mathbb R^2$ on the momentum space yields $\mathcal{BZ}=\mathbb R^2/\Lambda^\ast\cong \mathbb T^2$ for the Brillouin zone via identification. All 2D periodic (translational invariant) lattices $\Lambda$ yield the same result and, $\mathcal{BZ}$ will have torus topology. The maps of investigation are therefore $f\colon \mathbb T^2\to \mathcal M$. The induced maps on the level of homology and cohomology are given by
\begin{align}
f_{*}&\colon H_k(\mathbb T^2;\mathbb R)\to H_k(\mathcal M;\mathbb R), \label{hom_map}\\
f^{*}&\colon H^k(\mathcal M;\mathbb R) \to H^k(\mathbb T^2;\mathbb R). \label{hom_pullbmap}
\end{align}
$H_k$, $H^k$ denote the homology and cohomology groups of the spaces, respectively. $f_{*}$, $f^*$ are the corresponding (dual) homomorphisms between the groups. $f^*$ is of the pullback map type. The isomorphism $H^k(\mathcal M;\mathbb R)\cong H_k(\mathcal M;\mathbb R)$ is guaranteed if the coset space $\mathcal M$ is compact. Computation of homology groups is facilitated provided that the space admits a cell decomposition ($CW$-complex). In particular, we calculate first over $\mathbb Z$: $H_2(\mathbb T^2;\mathbb Z)=\mathbb Z$, $H_1(\mathbb T^2;\mathbb Z)=\mathbb Z\oplus \mathbb Z$. In general, replacing $\mathbb Z$ by $\mathbb R$ should be done with care, since the torsion subgroup vanishes in this operation, while the free part survives. For cases dealing with de Rham cohomology only the free abelian subgroup is relevant.

\subparagraph{Pull back bundles \& Bloch bundles.}
The map $\textbf{k} \rightarrow \hat H(\textbf{k})$ implies another relevant point: if the space $\mathcal M$ has a fibre bundle, then one can construct a fibre bundle over $\mathbb T^2$ with the same fibres; this is known as a \emph{Bloch bundle}. This follows from a general result: Let $f\colon \mathcal N\to \mathcal M$ be a map between manifolds and assume $\mathcal M$ admits a fibre bundle structure $(E,\pi,\mathcal M,F)$, $\pi\colon E\to \mathcal M$, then there exists a pull back bundle $f^{\bullet}E$ over $\mathcal N$, which has the same fibre $F$ (Steenrod \cite{Steen, Huse}).

\begin{figure}[H]
\centering
\begin{diagram}
f^{\bullet}E     &       \rTo^{pr_2}           &        E               \\
   \dTo^{pr_1}   &                             &       \dTo_{\pi}        \\
    \mathcal N   &        \rTo^{f}             &       \mathcal M
\end{diagram}
\caption{Induction of a pullback bundle $f^{\bullet}E$ over $\mathcal N$ from $f\colon\mathcal N\to\mathcal M$. The diagram is commutative, $\pi\circ pr_2=f\circ pr_1$.}
\label{fig:pullback_bundle}
\end{figure}
If we have two homotopic maps $f\sim g\colon \mathcal N\to\mathcal M$, then the induced bundles $f^\bullet E$ and $g^\bullet E$ will be equivalent \cite{Steen}.
We can combine this statement with a result about cell complexes \cite{Whitehead}:
If $\mathcal M,\mathcal N$ are manifolds which admit a cell decomposition and $f\colon \mathcal M\to \mathcal N$ is a map, then there exists a homotopic cellular map $g$ such that
\begin{equation}
f\sim g\colon \mathcal M\to \mathcal N,  \quad g(X^n)\subseteq Y^n, \label{cell_map}
\end{equation}
where $X^n, Y^n$ are $n$-skeletons - these are unions of cells\footnote{A cell $e^n$ is homeomorphic to $\mathbb R^n$ and to an open n-ball; $e^n\approx \mathbb R^n\approx  \mathbb D^{\circ n}$.} up to dimension $n$, e.g.
\begin{equation}
X^n=\bigcup_{\left\{e|\dim e\leq n\right\}} e,
\end{equation}
where $X^0\subseteq X^1\subseteq X^2\subset\cdots\subseteq X^n$. The general topological method which underlies the skeleton construction is:
\begin{remark} 
(Attaching spaces by maps) 
Let $X_0\subset X$ and $f\colon X_0\to Y$ be a map between topological spaces. Then we can create a new topological space $Y\bigcup_f X$ by the attaching map. The space $Y$ is naturally embedded into $Y\bigcup_f X$ through injection, i.e. $Y\hookrightarrow Y\bigcup_f X$. $Y\bigcup_f X:=X + Y/\sim$, $x_0\sim f(x_0)$ by identification.
\end{remark}
Attaching an $n$-cell $e^n$ to a (path-connected) space $Y$ to get $Y\cup e^n$ is given by $f\colon S^{n-1}\to Y$, where $S^{n-1}\approx \partial e^n$ (boundary of $n$-cell is ($n-1$)-dimensional sphere which is glued to space $Y$ via the mapping). 
This facilitates the understanding of Bloch bundles to some extent since $f$ needs to be described up to homotopy. An illustration is given by the following 
\begin{example}
($\mathbb T^2\rightarrow \mathbb S^2=\mathbb CP^1$)
The torus $\mathbb T^2=\mathbb S^1\times \mathbb S^1$ has the cell decomposition: one 0-cell $e^0$, two 1-cells $\left\{e^1_1=e^0\times e^1,e^1_2=e^1\times e^0\right\}$ and one 2-cell $e^2=e^1\times e^1$. $\mathbb S^2$ has an even simpler decomposition: one 0-cell $c^0$ and one 2-cell $c^2$. By the previous discussion $f\colon \mathbb T^2\to \mathbb S^2$ is homotopic to a cellular map $g$, and we must have
\begin{equation*}
g(e^0)=c^0, \quad g(e^1_1)=g(e^1_2)=c^0,
\end{equation*}
since the 1-skeleton of $\mathbb S^2$ is a one point-set according to the cell decomposition. The decision whether $f$ is topologically trivial or not depends on $g(e^2)$. If $g(e^2)=c^0$, then it is definitely a trivial map. However, non-trivial mappings arise if not all points of the 2-cell of the torus are mapped onto the 0-cell of the sphere.
\end{example}

\subparagraph{Topological properties of Berry Curvature.}

Assume the underlying parameter space of the Hamiltonian $\hat H$ is a compact manifold $\mathcal M$ of dimension $n$. The local parametrization on $\mathcal M$ shall be given by $\textbf{R}=(R_1,\dots,R_n)\in \mathbb R^n$, and Hilbert states are denoted by $\ket{\Psi(\textbf{R})}$; these are eigenstates of the problem $\hat H(\textbf{R})\ket{\Psi(\textbf{R})}=E(\textbf{R})\ket{\Psi(\textbf{R})}$. The gauge field (Berry connection) is given by 
\begin{equation}
\mathcal A(\textbf{R})=i\braket{\Psi(\textbf{R})|\nabla_{\textbf{R}}\Psi(\textbf{R})}.
\end{equation}
The (abelian) Berry curvature $\mathcal F\in \mathfrak u(1)\otimes \Omega^2(\mathcal M)$ assigned to the eigenstate over manifold $\mathcal M$ is obtained from the exterior derivative $\mathcal F=d\mathcal A$, $\mathcal F_{\nu\mu}=\partial_\nu\mathcal A_\mu - \partial_\mu\mathcal A_\nu$. The equation of motion for the gauge field $\mathcal A$ follows from $d\mathcal F=d^2\mathcal A\equiv 0$. $\mathcal F$ is considered a closed 2-form, i.e. $\mathcal [F]\in H^2(\mathcal M)$ (2nd cohomology group). Note that equation $\mathcal F=d\mathcal A$ has to be read as a local version and does not generally apply as a global condition on $\mathcal M$. Thus, if $\mathcal M$ has non-trivial homology, the cohomology class $[\mathcal F]$ is generally non-zero and implies topologically non-trivial effects, such as non-vanishing Chern numbers which can support topological phases. The pullback map in \eqref{hom_pullbmap} generates a Berry curvature $f^\ast \mathcal F$, which is associated with the Bloch bundle over $\mathbb T^2$, in particular $f^\ast [\mathcal F]\in H^2(\mathbb T^2)$. Since $H^2(\mathbb T^2;\mathbb R)=\mathbb R$ this curvature is generally non-trivial, but could turn out to be trivial, if $f$ is null-homotopic.

\subparagraph{Other topologies than $\mathbb T^2$?} 
First of all, one should refer to such topologies as \emph{non-standard} or \emph{atypical}. Recall that $\mathbb T^2$ has been obtained by a map $\mathbb R^2\rightarrow \mathbb R^2/\Lambda^\ast\cong \mathbb T^2$, where $\Lambda^\ast$ is a discrete (translational) subgroup of the full euclidean group $E(2)$, $\Lambda^\ast<E(2)$. The extension of the idea is to take other non-trivial discrete subgroups $G$ of $E(2)$ which provide us with an action $G\times \mathbb R^2\rightarrow \mathbb R^2$. This construction yields a specific tessellation of the plane. The search condition for $G$ is that the orbit space $\mathbb R^2/G$ should be compact, and $G$ is called a \emph{plane-crystallographic group} \cite{Coxeter}. Thus, together with the projection $\pi\colon\mathbb R^2\to\mathbb R^2/G$, the space $\mathbb R^2/G$ becomes a compact, connected 2D-manifold. Combinatorial topology provides us with the following classification: A compact, connected and closed 2-dimensional manifold $\mathcal S$ with genus $g\geq 1$ is the connected sum of either tori $\mathbb T^2$ or projective planes $\mathbb P^2$ (see \cite{Massey} for a proof), 
\begin{equation}
\mathcal S 
\cong \begin{cases}
\mathbb T^2\#\cdots \#\mathbb T^2 & \text{if orientable},     \\
\mathbb P^2\#\cdots \#\mathbb P^2  & \text{if non-orientable}.
\end{cases}
\end{equation}
 
\textsl{Gedankenexperiment}: Let Hamiltonian $\hat H$ be invariant under a plane crystallographic group $G$ such that $\mathbb R^2/G$ is non-orientable, for instance a projective plane $\mathbb P^2$ or Klein bottle $\mathbb P^2\#\mathbb P^2$. The underlying lattice can be thought to be generated by glide reflections, or similar group elements which are not discrete translations. Consider the abelian Berry curvature $[\mathcal F] \in H^2(\mathcal M)$, where $\mathcal M$ describes a coset space of the Hamiltonian, as given in \cref{fig:Altl_Zirn}. Due to $G$-symmetry, the Hamiltonian must be already uniquely determined on the fundamental region $\mathbb R^2/G$ (atypical $\mathcal{BZ}$). As an example assume $\mathbb R^2/G\cong\mathbb P^2$. Then, the corresponding map $f\colon \mathbb P^2\rightarrow \mathcal M$, $f\colon \textbf{k}\mapsto \hat H(\textbf{k})$ provides on the cohomological part $f^*\colon H^2(\mathcal M)\to H^2(\mathbb P^2)$. Here, we note that $H^2(\mathbb P^2;\mathbb R)=0$, as a consequence of being a non-orientable surface. Hence, $f^*\mathcal F$ must be globally trivial, i.e. $f^*\mathcal F=d\mathcal A$ on all $\mathbb P^2$, and consequently all (first) Chern numbers $\mathcal C$ vanish: $\mathcal C=\frac{1}{2\pi}\int_{\mathbb P^2}f^*\mathcal F=\frac{1}{2\pi}\int_{\mathbb P^2}d\mathcal A=0$. The topology of the band structure appears to be trivial, since a global potential exists. This extends to all non-orientable surfaces, in particular to the Klein bottle $\mathbb P^2\#\mathbb P^2$. 

This pertinent result is somehow counter-intuitive at first sight: although non-orientable surfaces (projective space or Klein bottle etc.) are topologically twisted, indicated by torsion elements in $H_1$, their 2nd degree topological structure eradicates the possibility for non-zero Chern numbers due to $H_2=0$. Hence, even if we manage to design artificial lattice systems with non-standard plane crystallographic group $G$ in the laboratory, measurements are predicted to yield zero Chern numbers in such systems.

\newpage

\section{Polaritonic Graphs}

\subsection{Scheme and Theory} 

The set-up scheme we use is based on joint optical microcavity pillars of polariton condensates arranged in the form of a 2D lattice configuration. Within the single-particle picture, and, by accounting for dominant transition amplitudes one can introduce corresponding hopping terms between sites of the lattice, such that the Hamiltonian will capture the relevant topological characteristics of this hybrid light-matter platform to a desirable degree. Although interesting exotic phases and highly non-linear excitation modes may arise in the presence of on-site polariton-polariton interactions, we omit these two-body operators in the following description. We will show how the existence of longitudinal (L) and transverse (T) polarization modes gives rise to Zeeman splitting by coupling to a magnetic field, and also cross-polarized terms. In other words, we will observe the way in which polarization modes can play a similar role as the usual spin degrees in electronic systems. 

\subsubsection{The Model and Hamiltonian} 
The starting form for the Hamiltonian of the tight-binding model can be written as (see chap. 3 \cite{Rips})
\begin{equation}
\hat H=\sum_{i,p}U_p\hat a_{p,i}^{\dagger}\hat a_{p,i} + \sum_{\braket{ij},p,p'}\bra{p,i}\mathcal J\ket{p',j}\hat a_{p,i}^{\dagger}\hat a_{p',j} + h.c., \label{Hamilton_polar1}
\end{equation}
subscripts $i,j$ run over lattice sites and $p\in\{L,T\}$ denotes the linear polarization mode of micro-cavity polaritons. The on-site potential is given by $U_p$. Operator $\mathcal J$ induces the nearest-neighbour hopping process between sites $i, j$. The creation and annihilation operators of cavity polaritons are denoted by $\hat a_{p,i}^{\dagger}$ and $\hat a_{p,i}$, respectively. The Bose-Einstein statistics of polaritons is encoded in the commutation relations 
\begin{equation}
[\hat a_{m}, \hat a_{n}]=[\hat a_{m}^{\dagger},\hat a_{n}^{\dagger}]=0, \quad [\hat a_{m},\hat a_{n}^{\dagger}]=\delta_{mn}.
\end{equation}

\subsubsection{Transformation of polarization modes in $\mathfrak H_2$}
The local on-site two-dimensional Hilbert space $\mathfrak H_2$ is spanned by the longitudinal and transversal polarization modes $\{\ket{L},\ket{T}\}$ of the polaritons. The structure of the on-site potential term requires us to transform this standard basis into the circular basis due to a Zeeman type of coupling of the magnetic field. As inferred from QED practice, the longitudinal-transverse basis can be mapped into the circular basis via 
\begin{equation}
\frac{1}{\sqrt{2}}(\ket{L}\pm i\ket{T})=\ket{\pm}.
\end{equation}
On the level of the annihilation operators, the inverse transformation yields
\begin{align}
\hat a_{L,j}&=\frac{1}{\sqrt{2}}(\hat a_{+,j}+\hat a_{-,j}) \label{LT_basistrf1}  \\
\hat a_{T,j}&=\frac{i}{\sqrt{2}}(\hat a_{+,j}-\hat a_{-,j}) \label{LT_basistrf2}
\end{align} 
We arrange polariton microcavity pillars in a lattice configuration which allows for a hopping between nearest neighbour sites via junctions (for technical details see \cite{Klembt,Nalit}). 
For (nn)-sites $\braket{ij}$ we find only diagonal terms in the $\{L,T\}$-basis, i.e. $\bra{L,i}\mathcal J\ket{L,j}=-t-\delta t$ and $\bra{T,i}\mathcal J\ket{T,j}=-t+\delta t$ ($t,\delta t>0)$. These results are strongly supported by a careful numerical analysis of the spinor component polariton wavefunction (methods \cite{Nalit}). The parameter $\delta t$ is due to the presence of TE-TM splitting in a junction connecting two sites.
From this observation and \cref{LT_basistrf1}, \cref{LT_basistrf2}, we only need the following expressions
\begin{align}
\hat a_{L,i}^{\dagger}\hat a_{L,j}&=\frac{1}{2}(\hat a_{+,i}^{\dagger}\hat a_{+,j}+ \hat a_{-,i}^{\dagger}\hat a_{-,j}+\hat a_{-,i}^{\dagger}\hat a_{+,j}+\hat a_{+,i}^{\dagger}\hat a_{-,j}),\\
\hat a_{T,i}^{\dagger}\hat a_{T,j}&=\frac{1}{2}(\hat a_{+,i}^{\dagger}\hat a_{+,j}+ \hat a_{-,i}^{\dagger}\hat a_{-,j}-\hat a_{-,i}^{\dagger}\hat a_{+,j}-\hat a_{+,i}^{\dagger}\hat a_{-,j}).
\end{align}

\paragraph{Peierls-phase free Hamiltonian.} 
We insert the above equations into \cref{Hamilton_polar1} 
\begin{equation}
\begin{split}
\hat H&=\mathcal B\sum_{i,\sigma}\sigma\hat a_{\sigma,i}^{\dagger}\hat a_{\sigma,i}-t\sum_{\braket{ij}}(\hat a_{L,i}^{\dagger}\hat a_{L,j}+\hat a_{T,i}^{\dagger}\hat a_{T,j} +h.c.)-\delta t\sum_{\braket{ij}}(\hat a_{L,i}^{\dagger}\hat a_{L,j}-\hat a_{T,i}^{\dagger}\hat a_{T,j} +h.c.)\\
&=\mathcal B\sum_{i,\sigma}\sigma\hat a_{\sigma,i}^{\dagger}\hat a_{\sigma,i}-t\sum_{\braket{ij},\sigma}(\hat a_{\sigma,i}^{\dagger}\hat a_{\sigma,j}+h.c.)-\delta t\sum_{\braket{ij}}(\hat a_{+,i}^{\dagger}\hat a_{-,j}+\hat a_{-,i}^{\dagger}\hat a_{+,j}+h.c.). \label{nophase_pham}
\end{split}
\end{equation}
The first part results from Zeeman-splitting, the second term resembles (nn)-hopping and the last sum consists of cross-polarized terms from the TE-TM splitting process. 

\paragraph{Geometrical origin of Peierls type phases and complete Hamiltonian.} 
So far, we have neglected Peierls-type phases $\exp(i\theta_{ij})$. We now show that these type of phases exist in principle between cross-polarized polaritons of neighbouring sites by the transformation of a well defined unitary operator on the polarization sector of the Hilbert space. Let $\mathcal{\hat U}(\theta)=\exp(-i\theta\hat \sigma_y)$ be the rotation operator acting on circular polarization states $\ket{\sigma}$, $\sigma=\pm$. Then,
\begin{gather}
\hat\sigma_y\ket{\sigma}=\sigma\ket{\sigma}, \\
\mathcal{\hat U}(\theta)\ket{\sigma}=e^{-i\theta\sigma}\ket{\sigma}.
\end{gather}
Let $\theta_{ij}\equiv \theta$ be the angle specifying the direction of the vector which connects the sites $i,j$ along the junction. Consider the adjoint operation of the operator $\mathcal{\hat U}$ on elements forming $\hat H\in End(\mathfrak H)$, i.e. on the space $End(\mathfrak H)\cong \mathfrak H\otimes\mathfrak H^*$:
\begin{align}
\ket{\sigma,i}\bra{\sigma',j}\longrightarrow \mathcal{\hat U}(\theta)\ket{\sigma,i}\bra{\sigma',j}\mathcal{\hat U}^{\dagger}(\theta)
&=e^{-i\theta\sigma}e^{i\theta\sigma'}\ket{\sigma,i}\bra{\sigma',j}\\
&=\begin{cases}
\ket{\sigma,i}\bra{\sigma,j} & \text{if $\sigma=\sigma'$},\\
e^{\pm2i\theta}\ket{\sigma,i}\bra{\sigma',j} & \text{if $\sigma\neq\sigma'$}.
\end{cases}
\end{align}
We see that non-trivial phases arise between cross-polarized polaritons in a 2D lattice configuration, i.e. we have transformations
\begin{equation}
\hat a^{\dagger}_{\sigma,i}\hat a_{\sigma',j}\longrightarrow e^{-i\theta(\sigma-\sigma')}\hat a^{\dagger}_{\sigma,i}\hat a_{\sigma',j}.
\end{equation}
To account for this, we derive for the \emph{full Hamiltonian} 
\begin{equation}
\hat H=\mathcal B\sum_{i,\sigma}\sigma\hat a_{\sigma,i}^{\dagger}\hat a_{\sigma,i}-t\sum_{\braket{ij},\sigma}(\hat a_{\sigma,i}^{\dagger}\hat a_{\sigma,j}+h.c.)-\delta t\sum_{\braket{ij}}(e^{2i\theta_{ij}}\hat a_{+,i}^{\dagger}\hat a_{-,j}+e^{-2i\theta_{ij}}\hat a_{-,i}^{\dagger}\hat a_{+,j}+h.c.). \label{polar_fullhamilton}
\end{equation}
Within our scheme, non-trivial phases can only occur in at least dimension $D=2$. In strictly one-dimensional systems the Hamiltonian will have the reduced form \cref{nophase_pham}. In other words, the phases are determined by the geometry of the 2-dimensional polaritonic lattice. 
The analysis reveals a purely geometrical origin of the phases; a mechanism which is slightly different than in systems considered by Peierls in his seminal work \cite{Peierls}. Recent advances even suggest the realization of Peierls-type phases (substitutions) in one-dimensional lattices of ultracold neutral atoms \cite{Jimenez_Spielman}.

\subsection{Computational Results} 

The hybrid light-matter structure of exciton-polaritons enables us to design and engineer synthetic Chern insulators. Two remarkable properties are: (1) The possibility of having a Zeeman splitting by the application of external magnetic fields \cite{Klembt}, (2) a spin-orbit type of coupling (SOC) due to TE-TM (transverse electric, transverse magnetic) splitting \cite{Panzarin, Sala}. A recent design of a polaritonic honeycomb lattice, related to Haldane's model, has been achieved in the work of Klembt \cite{Klembt}. Such a realization of Chern numbers $\mathcal C=\pm2$ has been predicted in the context of polaritonic topological insulators, utilizing graphene or honeycomb-like geometries \cite{Nalit}. By the bulk-boundary principle the non-trivial bulk topology corresponds to topologically protected edge states. A similar goal has been reached in the ultracold atom field: by loading atoms in an optical lattice of the honeycomb form, consisting of two triangular sub-lattices A/B with (nnn)-couplings, and then introducing laser-assisted nearest-neighbour tunnelling between the sub-lattices A and B \cite{Jotzu}. 

\subsubsection{Note on Symmetry Classes} 
\begin{figure}[h]
\centering
\subfloat[Kagome ground state level for $\mathcal B=0$, $\delta t=0$.]{{\includegraphics[width=.5\textwidth]{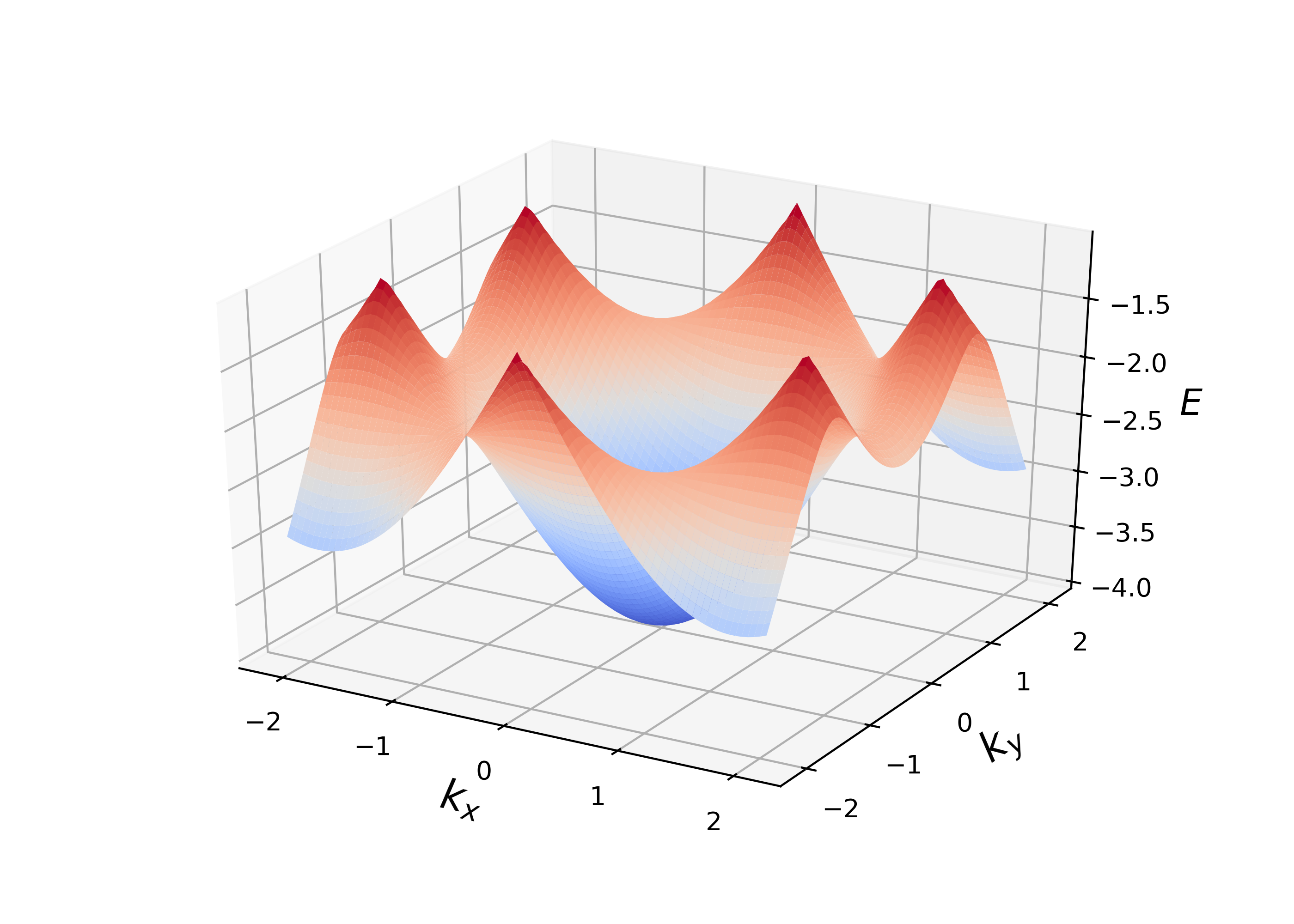}}}
\subfloat[Kagome band structure for $\mathcal B=0$, $\delta t=0$.]{{\includegraphics[width=.5\textwidth]{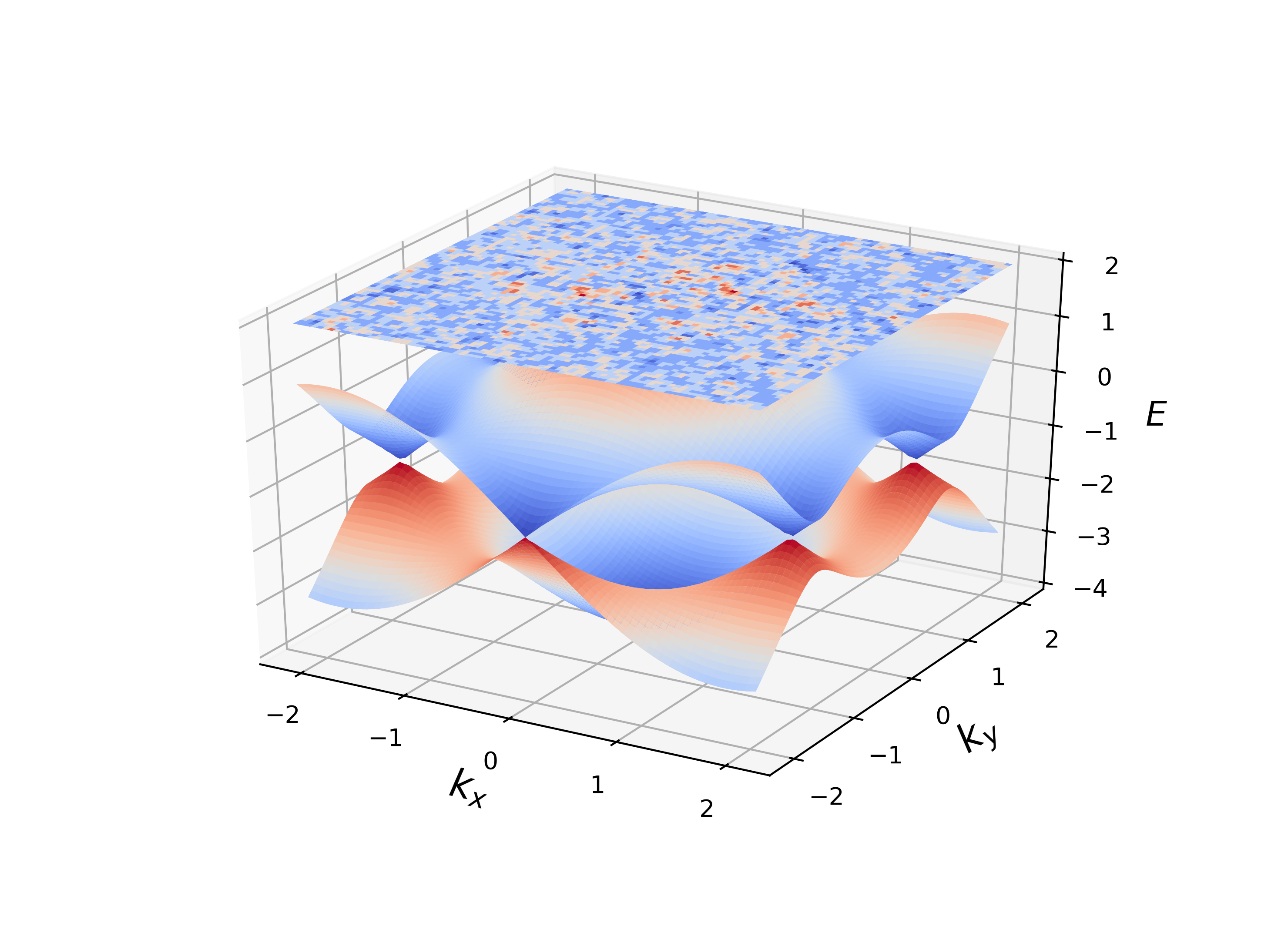}}}%
\caption{3D representation of the polaritonic Kagome band structure. Figure a) shows the lowest band and the symmetric arrangement of cones ($K$) in hexagonal form. In b) we see all three bands - the two lowest bands touch at appropriate Dirac points, while the highest band is completely flat.}
\label{fig:Kagome1}%
\end{figure}

The three classes of time-reversal (TR), particle-hole (PH) and chiral/sublattice (SL) symmetry have been used to classify the coset spaces of single particle Hamiltonians \cite{Altland} (\cref{fig:Altl_Zirn}). The topological structure of the coset space determines the existence of topological phases in a particular real dimension $D$ via the homotopy group $\pi_D$. One can show by direct comparison of $\pi_D$ of all coset spaces for $D=2$ that symmetry class $A$ Hamiltonians are of potential interest for engineering two-dimensional Chern insulators, characterized by a $\mathbb Z$-invariant. In these systems all three symmetries are absent. The existence of single positive flat bands with no negative counterparts in the band spectrum is a perfect indicator for PHS and SLS breaking. This follows directly from the discussion in appendix C. Although this is a strong condition, it is not the only possibility to determine absence of these symmetries within the band structure. The Kagome lattice with only nn-terms has such a flat band (see \cref{fig:Kagome1} b) and thus, is a suitable candidate for our modelling puposes. The breaking of time-reversal symmetry is achieved by coupling of cavity polarization modes to an externally applied magnetic field.


\subsubsection{Band Structures of the Kagome Lattice}

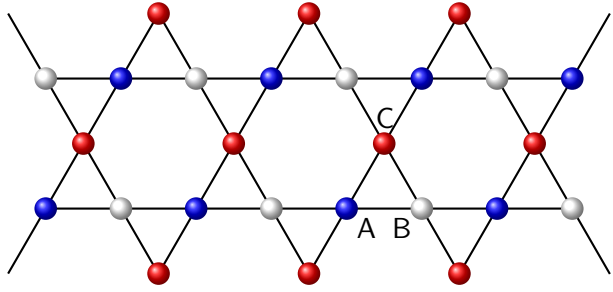
\begin{figure}[H]
\centering
\begin{tikzpicture}[scale=0.5] 
\tikzstyle{every node} = [ circle]
\path[draw,thick] (-4,0)--(2,0)--(10,0);
\path[draw,thick] (-4,3.46)--(2,3.46)--(10,3.46);
\path[draw,thick] (-5,5.19)--(-3,1.73)--(-1,-1.74);
\path[draw,thick] (-1,5.19)--(1,1.73)--(3,-1.74);
\path[draw,thick] (3,5.19)--(5,1.73)--(7,-1.74);
\path[draw,thick] (7,5.19)--(9,1.73)--(11,-1.74);
\path[draw,thick] (-5,-1.73)--(-3,1.73)--(-1,5.19);
\path[draw,thick] (-1,-1.73)--(1,1.73)--(3,5.19);
\path[draw,thick] (3,-1.73)--(5,1.73)--(7,5.19);
\path[draw,thick] (7,-1.73)--(9,1.73)--(11,5.19);
\foreach \b in {-2,0,...,5}
  \shade[shading=ball, ball color=blue](2*\b,0) circle (.3);
\foreach \w in {-1,1,...,5}
  \shade[shading=ball, ball color=white](2*\w,0) circle (.3);	
\foreach \w in {-2,0,...,5}
  \shade[shading=ball, ball color=white](2*\w,3.46) circle (.3);
\foreach \b in {-1,1,...,5}
  \shade[shading=ball, ball color=blue](2*\b,3.46) circle (.3);	
\foreach \r in {-1.5,0.5,...,5}
  \shade[shading=ball, ball color=red](2*\r,1.73) circle (.3);		
\foreach \ru in {-0.5,1.5,...,5}
  \shade[shading=ball, ball color=red](2*\ru,-1.73) circle (.3);			
\foreach \ro in {-0.5,1.5,...,5}
  \shade[shading=ball, ball color=red](2*\ro,5.19) circle (.3);	
\node[below right] (0) at (4,0) {\textsf A};
\node[below left] (1) at (6,0) {\textsf{B}};
\node[above right] (2) at (4.5,1.9) {\textsf{C}};
\end{tikzpicture}
\caption{Schematic of 2D polaritonic Kagome lattice structure (snapshot). Primitive unit cell vectors are $\textbf{d}_1=a(1,0)$ (\textsf{A}-\textsf{B}), $\textbf{d}_2=a(1/2,\sqrt{3}/2)$ (\textsf A-\textsf{C}). A unit cell is represented by the triangle $\triangle \textsf{ABC}$.}
\label{fig:polar_kagomelattice}
\end{figure}

The standard form of a Kagome lattice is given by (nn)-hopping processes and the unit cells consist of three sites or levels denoted by $A,B,C$ (\cref{fig:polar_kagomelattice}). We conclude from the periodic arrangement of the unit cells forming the lattice that the Fourier transformed Hamiltonian describes 3 bands over $\mathcal{BZ}$. However, the number of bands may be doubled due to existence of two polarization modes ($\sigma=\pm$), which are carried by each lattice site. With regards to the Kagome lattice, we get a total number of 6 bands, which makes the system potentially interesting for the purpose of exploring topological phenomena in hybrid materials, e.g. polaritonic topological insulators.

The translation vectors for shifting of unit cells are $\textbf{a}_i=2\textbf{d}_i$ (see \cref{fig:polar_kagomelattice}). Therefore, Kagome Bravais lattice unit cells are found at points $\textbf{R}(m,n)=m\textbf{a}_1+n\textbf{a}_2, m,n\in \mathbb Z$, with vectors $\textbf{a}_1, \textbf{a}_2$ being the generators of the lattice. 


\paragraph{Numerics.} 
The numerical computation of the energy bands is carried out with the Bloch Hamiltonian \cref{Bloch_kagome}, for which we set $a=1$ for the lattice geometry. For the detailed derivation of the Bloch Hamiltonian, we refer to the appendix.
\begin{equation}
\hat H(\textbf{k})=
\left(\begin{array}[pos]{ccc}
\mathcal B\hat\sigma_z & \mathcal M_{AB}(\textbf{k}) & \mathcal M_{AC}(\textbf{k}) \\
\mathcal M^{\dagger}_{AB}(\textbf{k}) & \mathcal B\hat\sigma_z & \mathcal M_{BC}(\textbf{k}) \\
\mathcal M^{\dagger}_{AC}(\textbf{k})  & \mathcal M^{\dagger}_{BC}(\textbf{k}) & \mathcal B\hat\sigma_z
\end{array}\right), 
\end{equation}
Furthermore, by scaling with respect to nearest-neighbour hopping amplitude $t=1$, we numerically solve the eigenvalue problem 
\begin{equation}
\hat H(\textbf{k})\ket{u_m(\textbf{k})}=E_m(\textbf{k})\ket{u_m(\textbf{k})}.
\end{equation}
The band structure is computed for different values of the magnetic field $\mathcal B$ and TE-TM splitting parameter $\delta t$. 
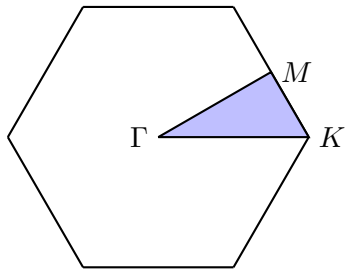
\begin{figure}[h]
\centering
\begin{tikzpicture}
 \foreach \i in {1,...,6}
   \draw[thick] (360/6*\i:2 cm) -- ({360/6*(\i+1)}:2 cm);
\node[left] (A) at (0,0) {$\Gamma$};
\node[right] (B) at (2cm,0) {$K$};
\node[right] (C) at (30:1.73cm) {$M$};
\draw[fill=blue!25, thick] (0,0) -- (2cm,0)-- (30:1.73cm)--(0,0);
\end{tikzpicture}
\caption{Schematic of the hexagonal $\mathcal{BZ}$ for the Kagome lattice. In accordance with the hexagonal shape, the computation of the energy bands can be performed along the closed path $\Gamma-K-M-\Gamma$ of high symmetry points.}
\label{fig:Hexag_BZ}
\end{figure}

\begin{figure}[h]
\centering
\subfloat[Kagome band structure  $\mathcal B=0.4$, $\delta t=0$.]{{\includegraphics[width=.5\textwidth]{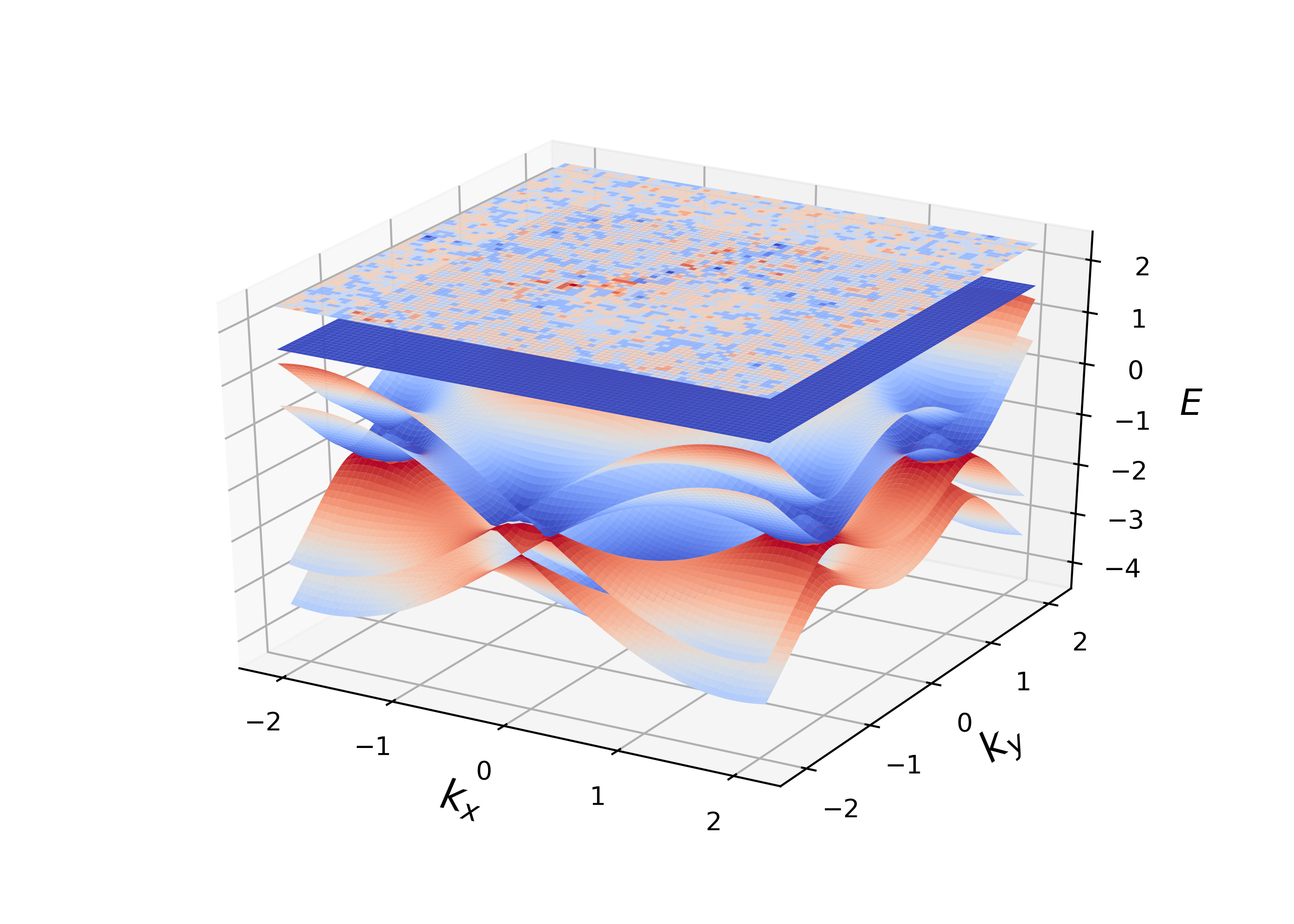}}} 
\subfloat[Kagome band structure for $\mathcal B=0.4$, $\delta t=0.2$.]{{\includegraphics[width=.5\textwidth]{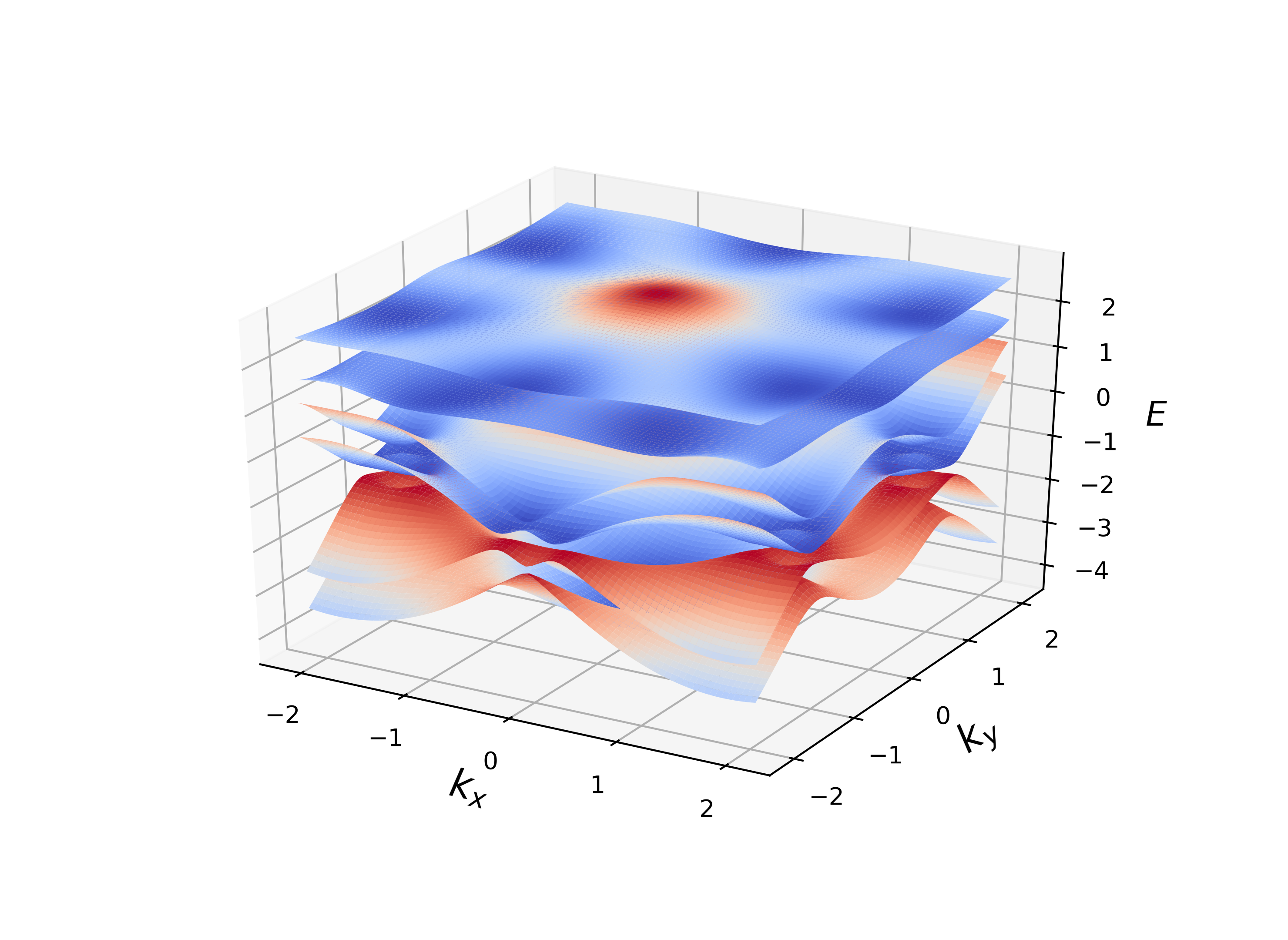}}}%
\caption{3D representation of the polaritonic Kagome band structure by including non-zero values for the magnetic field and the TE-TM splitting parameter. All 6 bands are displayed.}
\label{fig:Kagome2_3D}%
\end{figure}

\begin{figure}[h]
\centering
\subfloat[Kagome lattice bands for $\mathcal B=0.4$, $\delta t=0$.]{{\includegraphics[width=.5\textwidth]{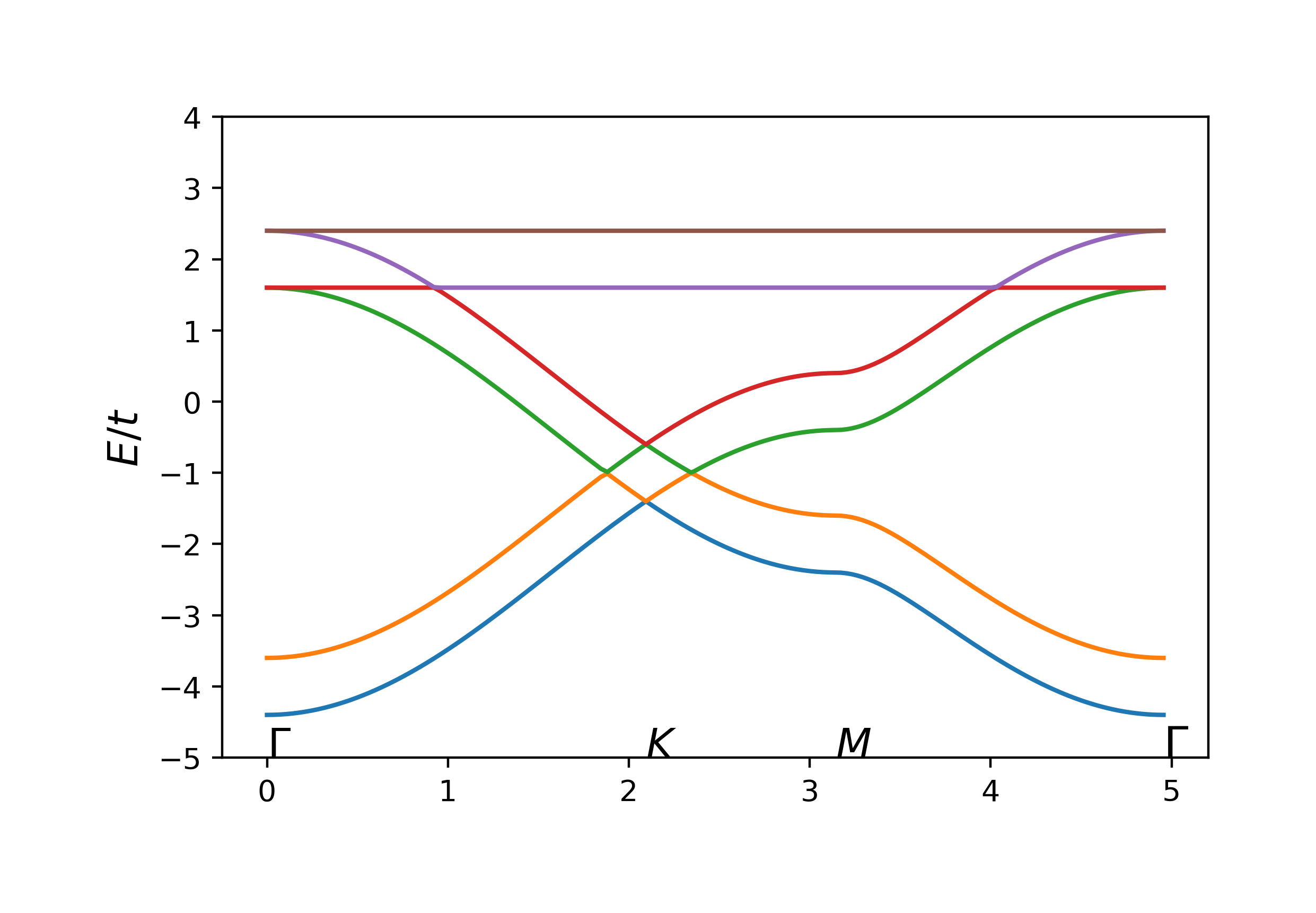}}}
\subfloat[Kagome lattice bands for $\mathcal B=0.4$, $\delta t=0.2$]{{\includegraphics[width=.5\textwidth]{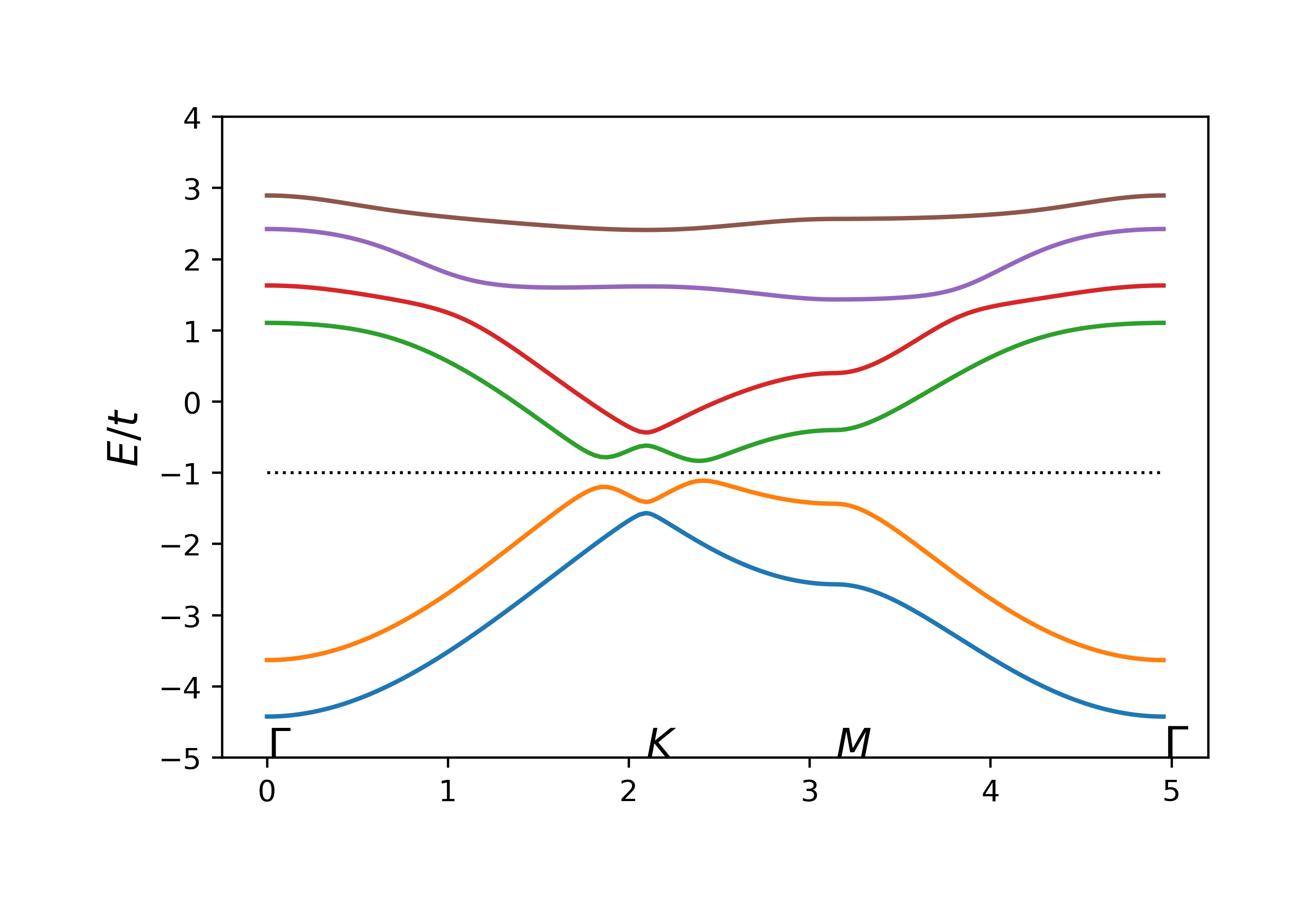}}}%
\caption{Gap opening mechanism in the polaritonic Kagome lattice by the interplay between the magnetic field $\mathcal B$ and the TE-TM splitting parameter $\delta t$. Crystallographic coordinates are used in $\mathcal{BZ}$ (\cref{fig:Hexag_BZ}), marking the points $\Gamma$ (centre of the hexagon), $K$ (Dirac cone), $M$. A gap opening mechanism is revealed around $E_g=-1$ (see b), clearly distinguishing between upper and lower bands.} %
\label{fig:Kagome2} %
\end{figure}

\begin{figure}[h]
\centering
\subfloat[Kagome lattice bands for $\mathcal B=0.4$, $\delta t=0.6$]{{\includegraphics[width=.5\textwidth]{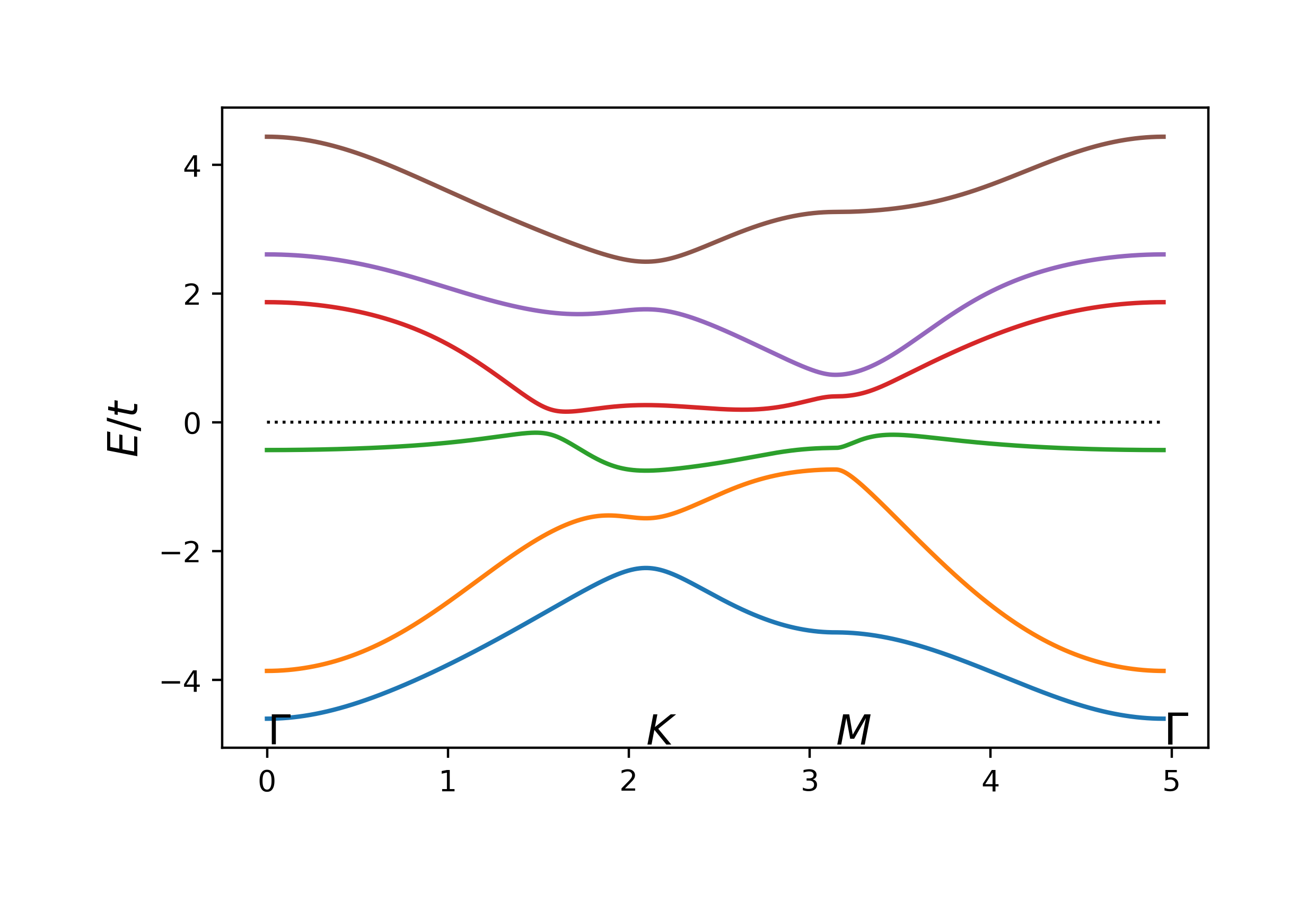}}}%
\subfloat[Kagome lattice bands for $\mathcal B=0.9$, $\delta t=0.6$]{{\includegraphics[width=.5\textwidth]{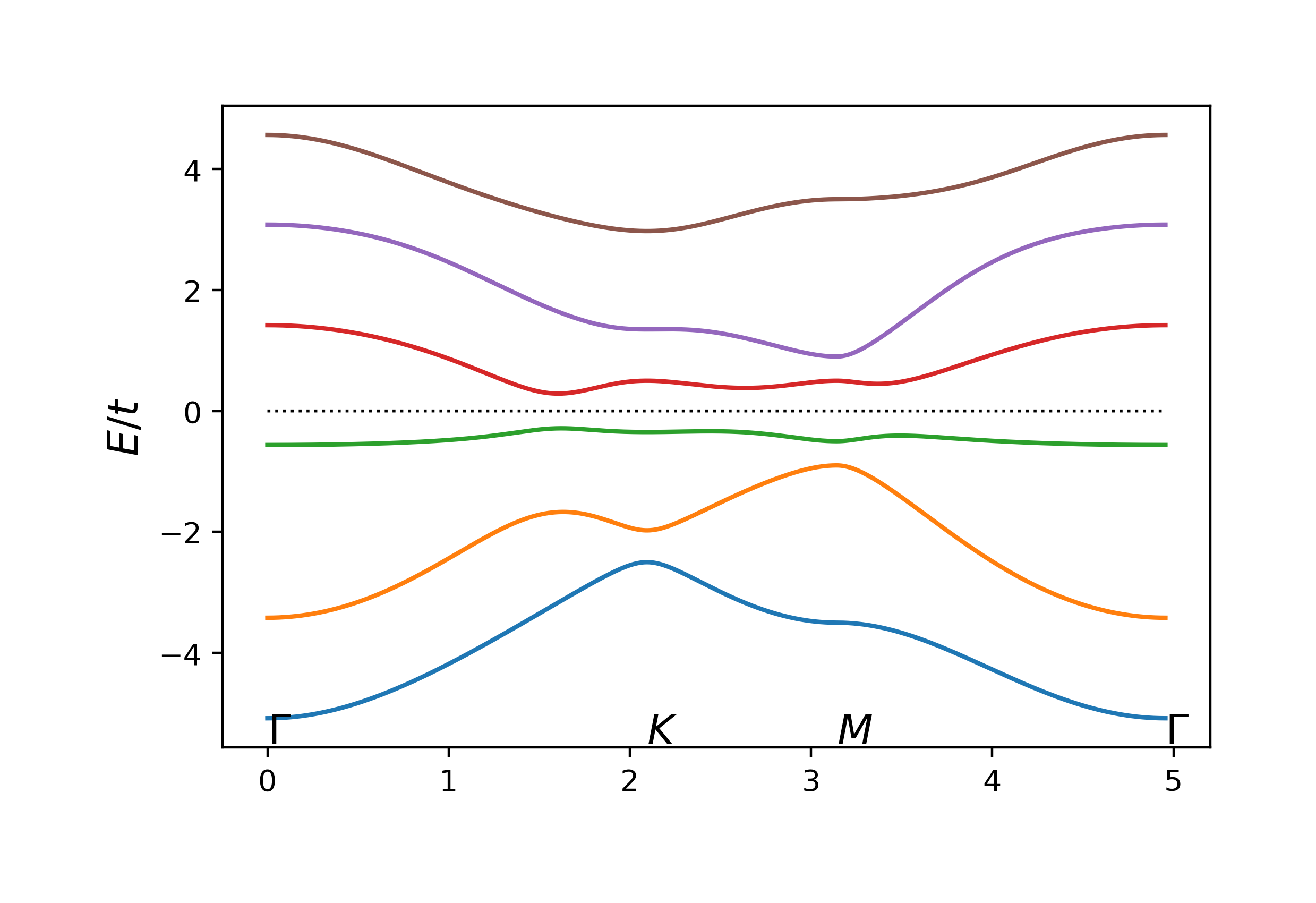}}}
\caption{Polaritonic Kagome lattice band structure for adjusting of the parameters to a) $\mathcal B=0.4$, $\delta t=0.6$, and to b) $\mathcal B=0.9$, $\delta t=0.6$, respectively. The diagrams display a potential gap around $E_g=0$. In spite of visual similarities, the Chern number distributions of the bands differ significantly, which is reported in the main text. Interestingly however, both systems belong to the same topological (valence band) insulator class.}%
\label{fig:Kagome3} %
\end{figure}

\vspace{5cm} 

\clearpage

The band structure of \cref{fig:Kagome2} b) suggests that there are $U(1)$-bundles over $\mathcal{BZ}$ for all the bands, since no band touching points exist. The band gap allows to calculate the Chern numbers for the two lowest bands below $E_g=-1$  
\begin{equation}
\mathcal C=\sum_{\left\{i:E_i<-1\right\}} \mathcal C^{(i)}=\frac{1}{2\pi}\sum_{\left\{i:E_i<-1\right\}}\int_{\mathcal{BZ}}\mathcal F^{(i)}=\frac{1}{2\pi}\sum_{\left\{i:E_i<-1\right\}}\int_{\mathbb T^2}\mathcal F^{(i)}(\textbf{k}), 
\end{equation}
where $\mathcal C^{(i)}$ denotes the first Chern number of the $i$-th band, $\mathcal{BZ}=\mathbb T^2$. It is the non--zero sum of Chern numbers of the populated bands below the gap which supports the topological insulating phase of the exciton-polariton crystal.

\begin{table}[h]
\centering
\begin{tabular}{l|l|l|l|l|l|l|l}
\hline
\hline
 \textsc{Parameters}  & \textsc{Band} :     & $E_1$    & $E_2$   & $E_3$   & $E_4$      & $E_5$  & $E_6$ \\
\hline
 $\mathcal B=0.4$, $\delta t=0.2$ &$\mathcal C^{(i)}$ num.: &  $-1.0(0)$   &    $3.0(0)$  &   $-2.0(0)$   &   $-1.9(9)$ & $2.9(9)$ & $-0.9(9)$ \\
\hline
 $\mathcal B=0.4$, $\delta t=0.6$ & $\mathcal C^{(i)}$ num.: &  $-1.0(0)$  & $3.0(0)$  &   $-2.0(0)$   &   $-2.0(0)$  &   $3.0(0)$ & $-0.9(9)$ \\
\hline
    & $\mathcal C^{(i)}$ theor.: &  $-1$ & $3$ & $-2$ & $-2$ & $3$ & $-1$ \\
\hline
\hline
$\mathcal B=0.9$, $\delta t=0.6$ & $\mathcal C^{(i)}$ num.: &  $-1.0(0)$   &    $0.0(0)$  &   $1.0(0)$   &   $1.0(0)$  & $-0.0(0)$  &  $-0.9(9)$  \\
\hline
                                 & $\mathcal C^{(i)}$ theor.: &  $-1$     &       $0$      & $1$         & $1$         & $0$         & $-1$  \\
\hline
\hline
\end{tabular}
\caption{Numerically extracted Chern numbers $\mathcal C^{(i)}$ for all six bands, for the cases in \cref{fig:Kagome2} b) and \cref{fig:Kagome3}. According to the algorithm CHN-AL, the computation has been performed using a discretization on a grid of $50\times 50$ sites over the Brillouin zone $[0,\pi]\times[-\frac{\pi}{\sqrt{3}},\frac{\pi}{\sqrt{3}}]$, where fluxes $\Omega_{nm}$ through each plaquette were obtained. The net flux gives the Chern number $\mathcal C^{(i)}=\frac{1}{2\pi}\sum_{n,m}\Omega^{(i)}_{nm}$ of the $i$-th band. The numerical accuracy is indicated within the brackets. Note that $\sum\mathcal C^{(i)}=0$, which is expected since the Berry curvatures of all bands sum up to zero (see appendix B.3).}
\label{fig:Chern_Kagome}
\end{table}
From \cref{fig:Chern_Kagome} and \cref{fig:Kagome2} b) we observe for $\mathcal B=0.4$, $\delta t=0.2$ the Chern number 
\begin{equation}
\mathcal C=\sum_{\left\{i:E_i<-1\right\}} \mathcal C^{(i)}=2,
\end{equation}
and this represents a $\mathcal C=2$-Chern insulator. On the other hand, we get a trivial Chern insulator ($\mathcal C=\sum_{\left\{i:E_i<0\right\}}\mathcal C^{(i)}=0$) for parameter values $\mathcal B=0.4$ and $\delta t=0.6$ (\cref{fig:Kagome3}).

\subsubsection{Discussion} 
The gap opening mechanism has been verified for parameter values $\delta t=0.2$, $B\in [0.3,0.5]$ by a numerical simulation. There is also a gap for adjusting the magnetic field to $\mathcal B=0.4$ and taking values $\delta t\in[0.2,0.45]$ - also, we note that no band degeneracies occur for these values which clearly supports the construction of $U(1)$-bundles assigned to each band.
From the numerical simulation we learn that the gap opening mechanism strongly depends on the interplay between the magnetic field and TE-TM splitting parameter. This parameter plays here an analogous role as the spin-orbit coupling (SOC) in electronic systems. 
Realistic parameter values (in units) are $\delta t=50-200\mu eV$, $\mathcal B=100-200\mu eV$ \cite{Nalit}. As a comparison, the gap size in polaritonic systems ($100\sim 150\mu eV$) exceeds the spin-orbit gap in graphene \cite{Yugui} by a factor of at least $10^2$. Driving the system into the $\mathcal C=0$-phase requires larger values for $\delta t$ - this is within reach as new results on cavity structures suggest \cite{Dufferwiel}.
Bulk-boundary correspondence predicts the existence of edge mode states at the interface of areas of differing Chern numbers. Propagation of these states is insensitive to perturbations and defects due to their topological origin. For the Kagome lattice, edge mode state analysis can be carried out by (non-resonant) excitation of polariton condensation into energy states in the band gap. The experimental realization should be analogous to Ref. \cite{Klembt}, but the theoretical study of the mode dynamics will be done elsewhere.  

\subsection{Spectral flattening method for topological bands} 

We consider a Bloch Hamiltonian $\hat H(\textbf{k})$ based on the assumption of translation invariance of the system in D dimensions and take $\ket{u_a(\textbf{k})}$ as corresponding Bloch states. For the following discussion no further symmetry conditions are imposed. The eigenvalue equation is given by
\begin{equation}
\hat H(\textbf{k})\ket{u_a(\textbf{k})}=E_a(\textbf{k})\ket{u_a(\textbf{k})}, \quad \textbf{k}\in \mathcal{BZ},
\end{equation}
where $a$ refers to the band number, $a\in \left\{1,\dots,n+m\right\}$. The ground state of a topological insulator is determined by the occupied bands below some gap within the band structure.
\begin{definition} (Projection operator $\mathbb {\hat P(\textbf{k})}$)
\begin{equation}
\mathbb {\hat P(\textbf{k})}:=\sum_{a=1}^m \ket{u_a(\textbf{k})}\bra{u_a(\textbf{k})},
\end{equation}
$m$ is the number of occupied (valence) bands.
\end{definition}
By construction, the projector is hermitian, $\mathbb{\hat P}^{\dagger}(\textbf{k})=\mathbb{\hat P}(\textbf{k})$, and idempotent, $\mathbb{\hat P}^2(\textbf{k})=\mathbb{\hat P}(\textbf{k})$.
\begin{definition} (Analytical index)
The analytical index of the projector $\mathbb{\hat P}(\textbf{k})$ is
\begin{equation}
\ind(\mathbb{\hat P}):=\dim\ker(\mathbb{\hat P})-\dim\ker(\mathbb{\hat P}^\bot),
\end{equation}
where $\mathbb{\hat P}^\bot$ is the orthogonal complement to $\mathbb{\hat P}$, i.e. $\mathbb{\hat P}^\bot=\hat{\mathds 1}-\mathbb{\hat P}$.
\end{definition}
The projector gives zero eigenvalues when being applied to unoccupied band states. These states are the zero modes of $\mathbb{\hat P}$ and span the kernel $\ker(\mathbb{\hat P})$. In the same way, occupied band states are the zero modes of $\mathbb{\hat P}^\bot$ and they span the kernel $\ker(\mathbb{\hat P}^\bot)$.
For convenience, we can re-define the operator:
\begin{definition} (involution operator $\mathbb {\hat O}(\textbf{k})$)
\begin{equation}
\mathbb {\hat O}(\textbf{k})= \hat{\mathds 1}-2\mathbb{\hat P}(\textbf{k}). \label{proj}
\end{equation}
\end{definition}
This new operator can be regarded as a \emph{spectral-flattened Hamiltonian}. $\mathbb {\hat O}(\textbf{k})$ has the properties:
\begin{enumerate}
\item $\mathbb {\hat O}^{\dagger}(\textbf{k})=\mathbb {\hat O}(\textbf{k})$ (hermitian)
\item $\mathbb {\hat O}(\textbf{k})\mathbb {\hat O}^{\dagger}(\textbf{k})=\mathbb {\hat O}^2(\textbf{k})=\hat{\mathds 1}$ (unitary)
\item $\Tr\left(\mathbb {\hat O}(\textbf{k})\right)=n-m$
\end{enumerate}
Note that $\dim\ker(\mathbb{\hat P}(\textbf{k}))=n$ and $\dim\ker(\mathbb{\hat P}^\bot(\textbf{k}))=m$. Hence, the index can be written as
\begin{equation}
\Tr\left(\mathbb {\hat O}(\textbf{k})\right)=n-m=\ind(\mathbb{\hat P}(\textbf{k})). \label{anal_index2}
\end{equation}
The class of all involution operators with the three given properties shall be denoted as $IU_{n,m}$. The operator $\mathbb {\hat O}$ is a unitary $(n+m)$-matrix with $m$ eigenvalues $\lambda=-1$ and $n$ eigenvalues $\lambda=1$, corresponding to filled and empty bands, respectively. Let $\mathbb {\hat O}\in IU_{n,m}$, $\mathcal U \in U(n+m)$ and regard the action
\begin{gather}
U(n+m)\times IU_{n,m}\rightarrow IU_{n,m},\\
(\mathcal U,\mathbb {\hat O})\mapsto \mathcal U\mathbb {\hat O}\mathcal U^{\dagger}. \label{conj_trf}
\end{gather}
One sees immediately that all 3 properties remain invariant under this transformation (conjugation operation). In particular, this is true for the analytical index \cref{anal_index2} under the conjugation operation \cref{conj_trf}. From a geometrical point of view we have generated an \textit{orbit} within $IU_{n,m}$. We show that this single orbit actually covers all the space $IU_{n,m}$.
\begin{lemma} 
$U(n+m)$ acts transitively on  $IU_{n,m}$ by conjugation operation in \cref{conj_trf}. In other words $IU_{n,m}$ is a homogeneous $U(n+m)$-space.
\end{lemma}
\begin{proof}
Let $\mathbb V\in IU_{n,m}$ be some involution operator. Then $\exists \mathcal W\in \mathbb U(n+m):$ 
\begin{equation*}
\mathcal W\mathbb V\mathcal W^{\dagger}=\diag(\lambda_1,\dots,\lambda_{n+m}),\quad \lambda_i=\pm 1.
\end{equation*}
Since $\Tr\left(\mathbb V\right)=n-m$ we can re-arrange the eigenvalues to get
\begin{equation*}
\mathbb D:=\diag(\underbrace{+1\dots,+1}_{n},\underbrace{-1,\dots,-1}_m)
\end{equation*}
This is achieved by using permutation matrices $\mathcal P_{\pi}$ ($\pi\in S_{n+m}$) which are orthogonal (hence unitary).
Combining all that together, we have $\mathbb V=\mathcal U\mathbb D\mathcal U^{\dagger}$ with the unitary operator $\mathcal U:=(\mathcal P_{\pi}\mathcal W)^{\dagger}$.
\end{proof}

From a group theoretical perspective it follows that there exists a bijection 
\begin{equation}
IU_{n,m}\stackrel{~}{\leftrightarrow} U(n+m)/I_{\mathbb D},
\end{equation}
where $I_{\mathbb D}$ is the isotropy group of $\mathbb D$ - this is defined by
\begin{equation}
I_{\mathbb D}=\{\mathcal U\in U(n+m)|\mathcal U\mathbb D=\mathbb D\mathcal U\}.
\end{equation}
It is straightforward to compute this isotropy group 
\begin{align}
I_{\mathbb D}&=U(n)\times U(m)\\
IU_{n,m}&\thickapprox U(n+m)/\left(U(n)\times U(m)\right) \label{grassmann1}
\end{align}
The equivalence in \cref{grassmann1} is more than just a bijection. The space $IU_{n,m}$ can be assigned a subspace topology, in particular Hausdorff space topology. Moreover, the action of $U(n+m)$ on $IU_{n,m}$ is continuous - this is true for the induced map $U(n+m)\rightarrow IU_{n,m}$ and for the inverse map, since $U(n+m)$ is compact and $IU_{n,m}$ is Hausdorff. For that reason, \cref{grassmann1} is a homeomorphism between topological spaces. This topological space is actually known as a \emph{Grassmannian} manifold $\mathbb G_{m,n+m}(\mathbb C)$ \footnote{Grassmannians are fundamental objects in pure geometry, and they have recently started to appear in the context of topological phases in condensed matter \cite{Altland, Kitaev}. Remarkably, they also seem to play a role for scattering amplitudes in particle physics.}, defined by  
\begin{equation}
\mathbb G_{m,n+m}(\mathbb C):= U(n+m)/\left(U(n)\times U(m)\right),
\end{equation}
which is compact and can be embedded into $\mathbb R^N$ for sufficiently large N due to Whitney's theorem. The dimension of the manifold $\mathbb G_{m,n+m}(\mathbb C)$ is 
\begin{equation}
\dim [U(n+m)/(U(n)\times U(m))]=(n+m)^2-(n^2+m^2)=2nm.
\end{equation}
Later on, we will specify the way in which the band numbers $m$ (valence bands) and $n$ (empty bands) determine the Bloch bundle structure of a Chern insulator in symmetry class A. For the rest of the discussion, we focus on symmetry class A Hamiltonians in real dimension $D$. In fact, the momentum space Hamiltonian has been transformed stepwise 
\begin{equation}
\hat H(\textbf{k})\longrightarrow \mathbb{\hat P}(\textbf{k})\longrightarrow \mathbb {\hat O}(\textbf{k}). \label{homot_deform}
\end{equation}
Operator $\mathbb {\hat O}$ is an equivalent replacement of the original Hamiltonian up to homotopic deformations. The outlined construction provides a map $f$
\begin{equation}
 f\colon \mathbb T^D\to U(n+m)/\left(U(n)\times U(m)\right),\quad \label{homtp_map}
 f\colon \textbf{k}\mapsto \mathbb {\hat O}(\textbf{k}),
\end{equation}
where the Brillouin zone $\mathcal{BZ}$ has been identified with the $D$-dimensional torus $\mathbb T^D$.
An \emph{adiabatic deformation} of the Hamiltonian $\hat H(\textbf{k})$ implies a continuous change of the parameters upon which the Hamiltonian depends, and we are naturally led to the homotopy class $[f]$ of \cref{homtp_map} - i.e. $[f]\in \{\mathbb T^D,\mathbb G_{m,n+m}(\mathbb C)\}$. However, it seems sufficient to understand homotopy classes $\{\mathbb S^{D},\mathbb G_{m,n+m}(\mathbb C)\}\cong \pi_{D}(\mathbb G_{m,n+m}(\mathbb C))$.
\begin{corollary}
The gapped topological phases in dimension $D$ are characterized by homotopy group $\pi_D(\mathbb G_{m,n+m}(\mathbb C))$. 
\end{corollary}
This corollary is an immediate consequence deduced from the classification of symmetry-protected gapped Hamiltonians by Altland and Zirnbauer in \cref{fig:Altl_Zirn}. It is important to understand how symmetry protection in terms of TRS, PHS and SLS alters the target space of the Hamiltonian. For that purpose, let us consider the target space of a system with sublattice symmetry:
\begin{example} 
(SLS protected Hamiltonian) 
Let $\hat H$ be only invariant under sublattice symmetry (SLS) $\hat{\mathcal S}$, $\hat{\mathcal S}\hat H\hat{\mathcal S}^{\dagger}=-\hat H$, see \cref{sub_latt_sym}. As in the case of the SSH model, we may perform an equivalence transformation on $\hat H$ which yields an off-diagonal block matrix form 
\begin{equation*}
\hat H(\textbf{k})\sim 
\left(\begin{array}[pos]{cc}
	0_n                  & \mathcal M(\textbf{k})\\
\mathcal M^\dagger(\textbf{k}) &        0_n 
\end{array}\right),
\end{equation*}
$\mathcal M(\textbf{k})\in U(n)$. Since $\pi_2(U(n))=0$, it is impossible to find topological insulators in dimension $D=2$ for  such SLS protected Hamiltonians. However, in real dimension $D=3$ we may find $\mathbb Z$-valued topological phases due to $\pi_3(U(n))=\pi_3(SU(2))=\mathbb Z$, $n\geq 2$.
\end{example}

\subsubsection{Application of Mapping Structure $\textbf{k}\longrightarrow \hat H(\textbf{k})\longrightarrow \hat{\mathbb P}(\textbf{k})\sim \hat{\mathbb O}(\textbf{k})$}

Consider the band structures in \cref{fig:Kagome2} b) and \cref{fig:Kagome3}, where we have different numbers of valence bands below the gap. With the help of the Altland-Zirnbauer symmetry classification \cref{fig:Altl_Zirn} we notice that $\hat H(\textbf{k})$ must represent a point in a Grassmannian, since its symmetry class is A. A topological insulator is determined by its valence Bloch bundle, or in other words, by the valence bands below the energy gap. For that we use the projector $\hat{\mathbb P}(\textbf{k})=\sum_a \ket{u_a(\textbf{k})}\bra{u_a(\textbf{k})}$ onto the subspace generated by eigenstates $\ket{u_a(\textbf{k})}$ corresponding to valence bands, or the equivalent spectral-flattened operator $\hat{\mathbb O}(\textbf{k})=\hat{\mathds 1}-2\mathbb{\hat P}(\textbf{k})$ \cref{proj}. Combining the numerical results with the spectral method, following scenarios can be deduced (\cref{fig:insulator_class_table}):

\begin{table}[h]
\centering
\begin{tabular}{l|l|l}
\hline
\hline
 $\mathcal B=0.4$, $\delta t=0.2$: & $\hat{\mathbb O}(\textbf{k})\in \mathbb G_{2,6}(\mathbb C)$ & $16$-\textsc{dimensional Grassmannian} \\
\hline
 $\mathcal B=0.4$, $\delta t=0.6$: & $\hat{\mathbb O}(\textbf{k})\in \mathbb G_{3,6}(\mathbb C)$ & $18$-\textsc{dimensional Grassmannian} \\
\hline
$\mathcal B=0.9$, $\delta t=0.6$: & $\hat{\mathbb O}(\textbf{k})\in \mathbb G_{3,6}(\mathbb C)$ &  $18$-\textsc{dimensional Grassmannian} \\                          
\hline
\hline
\end{tabular}
\caption{Different scenarios for topological insulators on Kagome lattice.}
\label{fig:insulator_class_table}
\end{table}

This means we have in fact a map
\begin{equation} 
f_{\delta t,\mathcal B}\colon \mathbb T^2\to \mathbb G_{m,6}(\mathbb C), \label{torus_grassm} 
\end{equation} 
depending on the experimental parameters $\delta t$, $\mathcal B$, and $m$ being the valence band number. Topological results of Milnor, Stasheff \cite{Milnor}, suggest a construction method of cell decompositions for Grassmann manifolds. In particular, one may use variants of Schubert calculus for verification. This mathematical tool gives us the right to construct a cellular map $g_{\delta t,\mathcal B}$, which is homotopic to $f_{\delta t,\mathcal B}$, as indicated in eq. \cref{cell_map},  
\begin{equation}
f_{\delta t,\mathcal B}\sim g_{\delta t,\mathcal B}\colon \mathbb T^2 \to X^2\hookrightarrow \mathbb G_{m,6}(\mathbb C),
\end{equation}
$X^2$ is the 2-skeleton of $\mathbb G_{m,6}$. This sets a restriction to the image of the torus $\mathbb T^2$: $g_{\delta t,\mathcal B}(\mathbb T^2)\subseteq X^2$. In addition, the existence of a monopole configuration in a Grassmannian yields the following interpretation (see \cref{fig:homgroup_Grass}): The two-dimensional Brillouin zone is mapped onto the 2-skeleton of the Grassmann manifold, and encloses the monopole in some non-trivial way. The observed non-zero Chern numbers $\mathcal C^{(i)}$ measure the net flux of the monopole field (expressed as Berry curvature $\mathcal F^{(i)}$ of a band), which then penetrates through the surface. 

\newpage 

\section{Advanced topological applications on the lattice}

\subsection{Exploiting Fibrations and Homotopy Sequences}

Firstly, I demonstrate the existence of a synthetic monopole in the coset space of symmetry class A Hamiltonians by the application of exact sequences of homotopy groups for fibre bundles or Serre fibrations. This perspective allows to supplement our findings made in the numerical analysis of the Kagome lattice: The Berry curvature can be regarded as generated by a monopole configuration in the parameter space of the Hamiltonian. Secondly, I provide a characterization of the valence Bloch bundles associated with the Chern insulators constructed on the Kagome lattice.

\begin{definition} 
(Stiefel manifold $\mathbb V_{k,n}(\mathbb C)$) 
The Stiefel manifold is defined as the set of ordered orthonormal $k$-frames in $\mathbb C^n$
\begin{equation}
\mathbb V_{k,n}(\mathbb C):=\left\{(e_1,\ldots,e_k)\in \mathbb C^{n\times k}|\braket{e_i|e_j}=\delta_{ij}\right\}
\end{equation}
\end{definition}
The topological structure of the Stiefel manifold is mathematically defined as a subspace topology.
\paragraph{First action on $\mathbb V_{k,n}$.}
The action on the Stiefel manifold is $U(n)\times \mathbb V_{k,n}(\mathbb C)\rightarrow \mathbb V_{k,n}(\mathbb C)$, $(\mathcal U,(e_1,\ldots,e_k))\mapsto (\mathcal  U e_1,\mathcal U e_2,\ldots,\mathcal U e_k)$, and it is transitive, which also applies to $SU(n)$. The isotropy group of the ordered tuple $(e_1,\ldots,e_k)$ is $U(n-k)$ (or $SU(n-k)$). Therefore, by the same token as given in the preceding paragraph, the manifold can be described as a homogeneous space 
\begin{equation}
\mathbb V_{k,n}(\mathbb C)\cong U(n)/U(n-k)\cong SU(n)/SU(n-k). \label{Stiefelm}
\end{equation}
\begin{remark}
Let $G$ be a Lie group and $H<G$ a closed subgroup, then $G/H$ will be the base manifold of a fibre bundle $(G,\pi,G/H,H)$ with the canonical projection $\pi\colon G\to G/H$ and fibre $H$. So, $H\stackrel{i}{\rightarrow}G\stackrel{\pi}{\rightarrow} G/H$. This well known  fact can be exploited for the computation of exact homotopy sequences.
\end{remark}
The corresponding long exact homotopy sequence for $\mathbb V_{k,n}$ is  
\begin{equation}
 \cdots\stackrel{\partial_*}{\longrightarrow}\pi_m(SU(n-k))\stackrel{i_*}{\longrightarrow} \pi_m(SU(n))\stackrel{\pi_*}{\longrightarrow} \mathbb \pi_m(\mathbb V_{k,n}(\mathbb C))\stackrel{\partial_*}{\longrightarrow} \pi_{m-1}(SU(n-k))\stackrel{i_*}{\longrightarrow}\cdots
\label{Stiefel_seq}
\end{equation}
We also need to compute some groups $\pi_m(U(n))$. First, it is known that spheres $\mathbb S^{2n-1}$ can be written as $\mathbb S^{2n-1}=U(n)/U(n-1)=SU(n)/SU(n-1)$ which constitutes a principal fibration with the exact sequences:
\begin{gather}
\cdots\stackrel{\partial_*}{\longrightarrow}\pi_m(U(n-1))\stackrel{i_*}{\longrightarrow} \pi_m(U(n))\stackrel{\pi_*}{\longrightarrow} \mathbb \pi_m(\mathbb S^{2n-1})\stackrel{\partial_*}{\longrightarrow} \pi_{m-1}(U(n-1))\stackrel{i_*}{\longrightarrow}\cdots \\
\cdots\stackrel{\partial_*}{\longrightarrow}\pi_m(SU(n-1))\stackrel{i_*}{\longrightarrow} \pi_m(SU(n))\stackrel{\pi_*}{\longrightarrow} \mathbb \pi_m(\mathbb S^{2n-1})\stackrel{\partial_*}{\longrightarrow} \pi_{m-1}(SU(n-1))\stackrel{i_*}{\longrightarrow}\cdots
\end{gather}
From the last two sequences we derive: $\pi_m(U(n))=\pi_{m}(U(n-1))$ and $\pi_m(SU(n))=\pi_{m}(SU(n-1))$ for $m<2n-2$, since in this case $\pi_{m+1}(\mathbb S^{2n-1})=\pi_{m}(\mathbb S^{2n-1})=0$. By induction we obtain some useful results: $\pi_0(U(n))=\pi_0(U(1))=0$ (connected Lie groups). $\pi_1(U(n))=\pi_1(U(2))=\pi_1(SU(2)\times U(1))=\pi_1(\mathbb S^3\times \mathbb S^1)=\pi_1(\mathbb S^1)=\mathbb Z \quad (n\geq 2)$. Also, $\pi_1(SU(n))=\pi_1(SU(2))=0$ and $\pi_2(SU(n))=\pi_2(SU(2))=\pi_2(\mathbb S^3)=0$. As a consequence from this and \eqref{Stiefel_seq}, we obtain for the lowest dimensional homotopy groups of the Stiefel manifold $\pi_0(\mathbb V_{k,n})=\pi_1(\mathbb V_{k,n})=\pi_2(\mathbb V_{k,n})=0, (n-k\geq 2)$.

\paragraph{Second action on $\mathbb V_{k,n}$.} We also have an action $U(k)\times \mathbb V_{k,n}(\mathbb C)\rightarrow \mathbb V_{k,n}(\mathbb C)$, $(\mathcal U,(e_1,\ldots,e_k))\mapsto (\mathcal U e_1,\mathcal U e_2,\ldots, \mathcal U e_k)$ where $\mathcal U$ is now a unitary $k\times k$-matrix. As before, the action is transitive, but the stabilizer is trivial, i.e. $\mathds 1_k$. Employing \eqref{Stiefelm} the resulting space is 
\begin{equation}
\mathbb V_{k,n}/U(k)\cong U(n)/(U(n-k)\times U(k))=\mathbb G_{k,n}(\mathbb C),
\end{equation}
which is a Grassmannian. 
We obtain the principal fibration $(\mathbb V_{k,n}(\mathbb C),\mathbb G_{k,n}(\mathbb C),U(k))$ for the Grassmann manifold with the projection $\pi \colon \mathbb V_{k,n}(\mathbb C)\to \mathbb G_{k,n}(\mathbb C)$. Correspondingly, the exact homotopy sequence is 
\begin{equation}
\cdots\stackrel{\partial_*}{\longrightarrow}\pi_m(U(k))\stackrel{i_*}{\longrightarrow} \pi_m(\mathbb V_{k,n}(\mathbb C))\stackrel{\pi_*}{\longrightarrow} \mathbb \pi_m(\mathbb G_{k,n}(\mathbb C))\stackrel{\partial_*}{\longrightarrow} \pi_{m-1}(U(k))\stackrel{i_*}{\longrightarrow}\cdots \label{Grassm_hom_seq}
\end{equation}
From all that we compute: $\pi_0(\mathbb G_{k,n}(\mathbb C))=\pi_1(\mathbb G_{k,n}(\mathbb C))=0$. Because of the vanishing homotopy groups of $\mathbb V_{k,n}$ for $d=1,2$, sequence \eqref{Grassm_hom_seq} yields the isomorphism $\pi_2(\mathbb G_{k,n}(\mathbb C))=\pi_1(U(k))=\mathbb Z$. 

\begin{table}[h]
\centering
\begin{tabular}{l|l|l}
\hline
\hline
$\pi_0(\mathbb G_{k,n}(\mathbb C))$ & $\pi_1(\mathbb G_{k,n}(\mathbb C))$ & $\pi_2(\mathbb G_{k,n}(\mathbb C))$ \\
\hline
$0 $       &  $0$            & $\mathbb Z$ \\
\hline
\hline
\end{tabular}
\caption{Summary of low-dimensional homotopy groups for the Grassmann manifold $\mathbb G_{k,n}(\mathbb C)$ (coset space of symmetry class A).}
\label{fig:homgroup_Grass}
\end{table}
This proves the existence of monopoles in the coset space of symmetry class A Hamiltonians. At the same time, we obtain a characterization of gapped phases of topological insulators for dimension $D=2$. In summary we have shown:
\begin{lemma}
Manifold $\mathbb G_{k,n}(\mathbb C)$ describes a polaritonic phase containing a monopole structure, and it admits a principal bundle of the type $(\mathbb V_{k,n}(\mathbb C),\mathbb G_{k,n}(\mathbb C),U(k))$ with respect to the band structure, in which case $k$ is the number of valence bands and $n$ is the total band number. $\square$
\end{lemma} 

\subsection{Hurewicz isomorphism, Bloch bundles \& index theorem}

\subsubsection{Monopole, Hurewicz's homomorphism and non-trivial Berry curvature}

There exists a remarkable relation between the artificial monopole configuration in the coset space, the aspherical property of $\mathbb G_{k,n}(\mathbb C)$ in dimensions $D<2$, and the resulting non-trivial curvature $\mathcal F$ assigned to the bands over the Brillouin zone $\mathbb T^2$. The link for this line of thought is:
\begin{remark}(Hurewicz's Theorem)
A topological space $X$ admits a homomorphism $\pi_k(X)\rightarrow H_k(X,\mathbb Z)$. If the space $X$ is aspherical for dimensions $i<k$, i.e. $\pi_i(X)=0$, then $\pi_k(X)\stackrel{\sim}{\rightarrow} H_k(X,\mathbb Z)$ is an isomorphism for $k\geq 2$.
\end{remark} 
Hurewicz's map $h$ can be constructed for any topological space $X$, and it has the following meaning: Given any spheroid $[s]\in \pi_k(X)$, i.e. a map $s\colon \mathbb S^k\to X$, we can assign a well-defined cycle in $H_k(X)$ given by  
\begin{equation}
h\left([s]\right):= s_*\left\{\mathbb S^k\right\}, \label{hurewicz_map}
\end{equation}
where $\left\{\mathbb S^k\right\}$ denotes the cohomology class in $H_k(\mathbb S^k)$ and $s_*\colon H_k(\mathbb S^k)\to H_k(X)$.
Putting together Hurewicz's theorem with the results of \cref{fig:homgroup_Grass} we conclude that $H_2(\mathbb G_{k,n}(\mathbb C))=\mathbb Z$ holds. One also infers $H^2(\mathbb G_{k,n}(\mathbb C);\mathbb R)\cong\mathbb R$ on the cohomological level. Since it is a one-generator group, all closed 2-forms, i.e. abelian Berry curvatures over the manifold, are effectively constructed from the single generator, denoted by $[\tilde{\mathcal F}]$. The map $f_{\delta t,\mathcal B}\colon \mathbb T^2\to \mathbb G_{m,m+n}(\mathbb C)$ gives rise to Berry curvatures for the various Bloch bands by the pull-back operation on the spectral level. The local form of the ith energy band is
\begin{equation}
f_{i,(\delta t,\mathcal B)}^\ast\tilde{\mathcal F}=\frac{1}{2}\tilde{\mathcal F}_{\mu\nu}\frac{\partial y_{(i)}^\mu}{\partial k^\alpha}\frac{\partial y_{(i)}^\nu}{\partial k^\beta}dk^{\alpha}\wedge dk^{\beta}, \label{der_B_curv}
\end{equation}
where a chart parametrization $y_{(i)}=y_{(i)}(\textbf{k})$ for $f_{i,(\delta t,\mathcal B)}$ on the torus manifold has been chosen in \cref{der_B_curv}. The abelian curvature components are given by $\tilde{\mathcal F}_{\mu\nu}=\partial_{\mu}\tilde{\mathcal A}_\nu - \partial_{\nu}\tilde{\mathcal A}_{\mu}$. 
Obstruction to choosing a global potential $\tilde{\mathcal A}$ for a band comes from the non-trivial cohomology group $H^2$ and \cref{der_B_curv}. Numerical evidence of Chern numbers in \cref{fig:Chern_Kagome} supports this result. Moreover, we observe topological robustness of the band structure in regions of the $(\delta t,\mathcal B)$-space where the Hamiltonian is gapped. Possible parameter adjustments are shown in the figures - along with corresponding Chern number distributions which arise in respective regions of the parameter space.

\subsubsection{Characterization of valence Bloch bundles}

We combine our numerical results on the Kagome lattice with the statement in \cref{fig:pullback_bundle} in order to describe the pullback bundles over the Brillouin zone $\mathbb T^2$.

\begin{figure}[H]
\centering
\subfloat[Trivial Bloch bundle for $\mathcal C=0$.]
{\begin{diagram}
\mathbb T^2\times U(3)\cong & f^\bullet_{0.6,0.4} \mathbb V_{3,6}(\mathbb C)       &   \rTo^{pr_2}  & \mathbb V_{3,6}(\mathbb C)   \\
                            &          \dTo^{pr_1}   &                            &          \dTo_{\pi}          \\
                            &         \mathbb T^2    &    \rTo_{f_{0.6,0.4}}      &    \mathbb G_{3,6}(\mathbb  C)
\end{diagram}}
\\
\subfloat[Non--trivial Bloch bundle for $\mathcal C=2$.]
{\begin{diagram}
 \mathbb T^2\times U(2)\neq   &f^{\bullet}_{0.2,0.4}\mathbb V_{2,6}(\mathbb C) &  \rTo^{pr_2}  & \mathbb V_{2,6}(\mathbb C)\\
            &\dTo^{pr_1}   &                             &       \dTo_{\pi}    \\
            &\mathbb T^2   &        \rTo_{f_{0.2,0.4}}   &       \mathbb G_{2,6}(\mathbb C)
\end{diagram}}
\caption{Pullback bundle diagrams for polaritonic Kagome configurations. The map is $f_{\delta t,\mathcal B}\colon \mathbb T^2\to \mathbb G_{m,6}(\mathbb C), \textbf{k}\mapsto \hat{\mathbb O}(\textbf{k})$, given in \cref{torus_grassm}. $\pi\circ pr_2=f_{\delta t,\mathcal B}\circ pr_1$. It is sufficient to describe $f_{\delta t,\mathcal B}$ up to homotopy, since homotopic maps yield equivalent pullback bundles.}
\label{fig:kagome_bloch_bundles}
\end{figure} 


\begin{table}[H]
\centering
\begin{tabular}{l|l|l}
\hline
\hline
                     & $\delta t=0.2$, $\mathcal B=0.4$  &  $\delta t=0.6$, $\mathcal B=0.4$  \\
\hline
 Grassmannian $\mathbb G_{m,n+m}(\mathbb C)$   &       $\mathbb G_{2,6}(\mathbb C)$  &        $\mathbb G_{3,6}(\mathbb C)$  \\
\hline
Fibre bundle (auxiliary) & $(\mathbb V_{2,6}(\mathbb C),\mathbb G_{2,6}(\mathbb C),U(2))$& $(\mathbb V_{3,6}(\mathbb C),\mathbb G_{3,6}(\mathbb C),U(3))$\\
\hline
 Valence Bloch bundle & $(f^{\bullet}_{0.2,0.4}(\mathbb V_{2,6}(\mathbb C)),\mathbb T^2,U(2))$ & $(\mathbb T^2\times U(3),\mathbb T^2, U(3))$\\
\hline
  Valence bands below gap        &    $2$      &           $3$             \\
\hline
  Chern number $\mathcal C$      &    $2$      &           $0$              \\
\hline
$N_L-N_R$                        &    $2$      &           $0$                \\
\hline 
 $\ind(\mathbb{\hat P})=n-m$     &    $2$      &           $0$                  \\
\hline
\hline 
\end{tabular}
\caption{Topological properties of polariton insulators on the Kagome lattice. $N_L/N_R$ denote the numbers of left/-right propagating edge mode states on the boundary. Relation $N_L-N_R=\mathcal C$ is known from the bulk-boundary correspondence. Principal bundles are written in standard notation as triplets.}
\label{fig:Summary_Kagome}
\end{table}
We make some comments on the results in \cref{fig:Summary_Kagome} and \cref{fig:kagome_bloch_bundles}: We observe that the valence Bloch bundle $(f^{\bullet}_{0.2,0.4}(\mathbb V_{2,6}(\mathbb C)),\mathbb T^2,U(2))$ cannot be trivial since $\mathcal C=2$, i.e. $f^{\bullet}_{0.2,0.4}(\mathbb V_{2,6}(\mathbb C))\neq \mathbb T^2\times U(2)$. The other diagram suggests the equivalence $\mathbb T^2\times U(3)\cong f^\bullet_{0.6,0.4} \mathbb V_{3,6}(\mathbb C)$. However, for both cases, the well known bulk-boundary correspondence predicts $N_L-N_R=\mathcal C$. Note that $\ind(\mathbb{\hat P})=n-m$ for $\mathbb G_{m,n+m}$ is a purely algebraic result.

\subsubsection{Index formula(s), implications and conjectures}

The above results indicate a remarkable relation which can be stated more rigorously:
\begin{observation}(Index Formula)  
Let $\hat H$ be a gapped symmetry class A Hamiltonian defined on some two-dimensional polaritonic lattice, let $E_g$ be the energy value within the spectral gap distinguishing between $m$ valence and $n$ unoccupied bands. Then, the index of the projector $\mathbb{\hat P}_{\mathbb G_{m,n+m}(\mathbb C)}$ assigned to the energy band structure is given by 
\begin{equation}
\frac{1}{2\pi}\int_{\mathds T^2} \sum_{\{i|E_i<E_g\} }\mathcal F^{(i)} =\ind\left(\mathbb{\hat P}_{\mathbb G_{m,n+m}(\mathbb C)}\right),
\label{main_index}
\end{equation} 
which represents an index theorem for the investigated class A Chern insulator. The Berry curvatures $\mathcal F^{(i)}$ are represented by abelian two-forms over the torus $\mathds T^2$, which can be obtained from spectral level pull-back maps. Each curvature can be replaced by its corresponding cohomology class $[\mathcal F^{(i)}]$ in the cohomology ring $H^*(\mathds T^2)$. 
\end{observation}

One insightful corollary can be inferred from classical degree theory. The argument goes as follows: The map $f\colon \mathds T^2 \to \mathbb G_{m,n+m}$, $\textbf{k}\mapsto \mathbb {\hat O}(\textbf{k})$ can be associated with spectral-level maps $f_i\colon \mathds T^2 \to \mathbb G_{m,n+m}$, which correspond to the energy levels $E_i(\textbf{k})$. As before, this gives rise to maps $f_{i,*}\colon H_2(\mathds T^2;\mathbb Z)\to H_2(\mathbb G_{m,n+m};\mathbb Z)$ and their dual maps $f_{i}^{*}\colon H^2(\mathbb G_{m,n+m};\mathbb R)\to H^2(\mathds T^2;\mathbb R)$. With regards to the homology groups, we know that $H_2(\mathds T^2;\mathbb Z)\cong H_2(\mathbb G_{m,n+m};\mathbb Z)\cong\mathbb Z$ holds, where both groups have exactly one generator, denoted by $[\mathds T^2]$ and $[\mathcal Y]$, respectively. Due to Stokes theorem, the integral in \cref{main_index} is independent of the homology class representative of the torus surface, and the cohomology class representative of the Berry curvature. Thus, we can rewrite it as $\int_{\mathds T^2}\mathcal F^{(i)} = ([\mathcal F^{(i)}], [\mathds T^2])$ in terms of classes, where we have used de Rham's pairing function $(\cdot,\cdot )\colon H^k(\mathds T^2)\times H_k(\mathds T^2) \to \mathbb R$ for $k=2$. Further, by choosing one representative for each Berry curvature,  we write the left-hand sum in \cref{main_index} as $\sum_i(\mathcal F^{(i)}, [\mathds T^2]) = \sum_i(f_{i}^{*}\tilde{\mathcal F}, [\mathds T^2]) = \sum_i(\tilde{\mathcal F}, f_{i,*}[\mathds T^2])$, where $\tilde{\mathcal F}$ is a suitable representative generator of $H^2(\mathbb G_{m,n+m};\mathbb R)$. From the above discussion, it follows that $f_{i,*}[\mathds T^2] =q_i\cdot [\mathcal Y]$ with a multiple $q_i\in\mathbb Z$ - according to homology theory, this integer is invariant under homotopy deformations of the map $f_i$, and it defines the degree of the map, $\deg f_i$. In a certain geometrical sense, the degree measures the wrapping number of the torus surface on the two-dimensional image space for the given map. Using the method of cell decompositions of spaces and cell maps, as introduced in \cref{cell_map}, we can deform each $f_i$ into a cellular map $g_i$ such that $f_i\sim g_i\colon g_i(\mathds T^2)\subseteq \mathds X^2$ holds, where $\mathds X^2$ is the two-skeleton of the Grassmann manifold. Due to homotopy invariance the degrees will be equal, i.e. $\deg f_i=\deg g_i$. Accordingly, the restriction of the image to the two-skeleton allows us to consider $g_{i,*}\colon H_2(\mathds T^2;\mathbb Z)\to H_2(\mathds X^2;\mathbb Z)$. As a result, we may 'disentangle' above formula in the following way
\begin{equation}
\left(\frac{1}{2\pi}\right)\left(\sum_{\{i|E_i<E_g\}}\deg f_i\right)\cdot \left([\tilde{\mathcal F}],[\mathcal Y]\right)=\ind\left(\mathbb{\hat P}_{\mathbb G_{m,n+m}(\mathbb C)}\right), \label{main_indexv2}
\end{equation}
where the Berry curvature $\tilde{\mathcal F}$ is integrated over some two-dimensional subspace of the Grassmannian, i.e. 
\begin{equation}
\left([\tilde{\mathcal F}],[\mathcal Y]\right) = \int_{\mathcal Y \subset \mathbb G_{m,n+m}} \tilde{\mathcal F}. 
\end{equation}
Note, that the sum in \cref{main_indexv2} runs over degrees of highly non-trivial maps, in contrast to \cref{main_index}, where the computation has been done by summing over individual Chern numbers. In practical applications, it is convenient to numerically compute each band Chern number via some algorithm, e.g. \cref{CHN_AL}.
The analytical computation of degrees of maps between arbitrary topological spaces is challenging, and according to my own knowledge a general approach does not yet exist.

More information on the structure of the representative Berry curvature $\tilde{\mathcal F}$ can be obtained by means of complex geometry. Under the assumption of the coset space being $\mathbb CP^n$, we identify it as a complex Kähler manifold, which naturally carries a Kähler-potential $\mathcal K$. From this potential, we can compute a metric and a non-trivial 2-form corresponding to Berry's curvature, thus directly reading out the degree assigned to the lowest valence band.

We would like to emphasize that the statements encoded in \cref{main_index} and \cref{main_indexv2} are far from being obvious. Why do the Chern numbers of bands below the gap sum up to the analytical index of $\mathbb{\hat P}$ for the manifold?
The mathematical structure of the conjecture is reminiscent of the \emph{Atiyah-Singer index theorem}. Recall that $\ind(\mathbb{\hat P})$ is the difference between zero eigenmodes of $\mathbb{\hat P}$ and the zero modes of $\mathbb{\hat P}^\bot$. From this point of view, the bulk-boundary correspondence must be interpreted as the manifestation of an index theorem. Attempts for a rigorous proof and development of the topic are deferred to future work. However, there exist already connections to K-theory and non-commutative geometry, and these had been worked out by Bellissard and co-workers \cite{Bellissard} - with the purpose of understanding the quantum hall effect.

At first sight, it may seem plausible that one may apply the index theorem to other lattice systems by considering only the band structure. However, this is not true as can be seen from the following pertinent counterexamples: 

1) (specific case) Consider the well known two-level system of Haldane, which has a Chern number $\mathcal C=1$ for the ground state. A blind application of the index theorem to this model would, however, imply a Chern number $\mathcal C=1-1=0$. The contradiction is due to the existence of several symmetries in Haldane's model, one of them is the preserved time-reversal symmetry (TRS). In the polaritonic lattice, TRS is explicitly broken by an external magnetic field. Moreover, the other two relevant symmetries (PHS and SLS) are absent in the polaritonic systems considered in this work. 

2) ('generalization') Without referring to a specific system, we can provide the following construction. Consider a Hamiltonian defined over some lattice, such that its parameter space or coset space $\mathcal M$ has a trivial second cohomology group $H^2(\mathcal M)=0$. Then, one may easily infer that all band Chern numbers over the Brillouin zone must be zero - no topological (insulator) material exists in this case, even if a fully gapped band structure is displayed. A more general topological argument reveals that even if the space $\mathcal M$ has non-trivial homotopy group $\pi_K(\mathcal M)$, it may happen to have a trivial homology group $H_K$ following from a surjective Hurewicz mapping $\pi_K(\mathcal M)\to H_K(\mathcal M)\cong H^K(\mathcal M)$ ($M$ compact), whose kernel trivializes the map. In the recent past, pure topologists have constructed such abstract instances of not necessarily homogeneous spaces that are acyclic in some dimension, but which do have non-trivial homotopy group in the same dimension. In other words, it shows that non-trivial spaces carrying high-dimensional defects or textures, can have trivial homology - therefore, homological operations are not sufficient to detect these objects, and one must rely on detection via homotopy groups. 

For symmetry class A in 2D, we have observed the special case that the second homology group $H_2$ is non-trivial due to Hurewicz mapping - the coset space is aspherical in dimensions $K<2$, but for $K=2$ it is non-trivial - hence, the second homotopy group is directly isomorphic to $H_2$, which itself is isomorphic to the cohomology group $H^2$. This cohomology group measures existing non-trivial Berry curvatures (co-cycles) over the space. In a more physical wording, the existence of a monopole defect in the coset space can give rise to potentially non-trivial (abelian) Berry curvatures connected with the bands of the energy spectrum. As demonstrated, this situation occurs for the studied Grassmannian coset space describing the polaritonic topological insulator.

\section{Conclusion and outlook}

Polaritonic lattices offer ideal platforms for nano-fabrication of etched 2D materials with topological properties, such as Chern insulators on lattice configurations. These lattices can be deployed as test beds for topological phase engineering in condensed matter and quantum physics, and for technological applications based on long range spatial coherence. The observed gap opening mechanism is not based on any spin-orbit coupling, as e.g. seen in electronic systems, but follows from competing values of the magnetic field and TE-TM splitting. The TE-TM and Zeeman splitting terms allow for controllable simulation of potential landscapes. On the Kagome lattice, we have demonstrated the existence of trivial and non-trivial Chern insulators in different regimes of the phase space. In order to gain deeper insight, I have re-formulated everything in the language of fibre bundles (Bloch bundles), homotopy, homology-cohomology duality, artificial gauge fields, and complex geometry. The investigation leads to a remarkable index theorem for symmetry class A Chern insulators, which shows close similarity to other index theorems, in particular the prominent Atiyah-Singer index. A rather delicate question is how to elegantly design Chern insulators in the form of two-dimensional single sheets, such that one can systematically manipulate the number $m$ of valence bands, and $n$ of unoccupied bands. This is equivalent to obtaining spectral-flattened Hamiltonians in $\mathbb G_{m,n+m}(\mathbb C)$, with tunable numbers $m$, $n$. Using the index theorem, we can partially resolve the problem: consider the simplest example, the projective space $\mathbb CP^n=\mathbb G_{1,n+1}(\mathbb C)$, which corresponds to a gapped system with one (occupied) valence band and $n$ empty bands. The Chern number would be $n-1$, implying that one can engineer insulators of arbitrary Chern numbers in that way - at least in principle by relying on the index formula. However, for the moment, it appears that we do not have a fully systematic approach and absolute control over engineering the targeted band spectrum. A partial theoretical consideration has been initiated by the proposal. Nevertheless, the investigation of various other models is straightforward: e.g., for a polaritonic Ruby lattice one shall find $6\times 2$ spectral bands. The SOC analogue gap opening mechanism as well as related topological properties can be investigated by strategies outlined in this work. The implementation could be advanced further by considering different junctions (with two or more hopping amplitudes $t_i$ and TE-TM splitting parameters $\delta t_i$) between the etched optical micro-cavity pillars. 

We have not touched upon the topic of aperiodic lattices, e.g. \emph{quasicrystals} based on the Ammann-Beenker tiling \cite{Loring2}. From a computational point of view, the lack of translational invariance sets some numerical challenges to the analysis of the bulk spectrum and the definition of topological real-space invariants. Nonetheless, in analogy to the concept of the Fredholm operator index, several authors have introduced and investigated a local and global index (Bott index) for this purpose \cite{Loring1, Toniolo}. Another complementary research direction could include the study of on-site polariton-polariton interaction effects, not necessarily in terms of topology, but in order to construct and understand quantum many-body ground states via simulation of large N-populations of polaritons on a lattice.

\paragraph{Acknowledgement.} 
K.R. is grateful for a sponsorship which enabled research in the CMT group at Lancaster University, UK, and would like to thank his supervisor J. Ruostekoski during the work at Lancaster for fruitful discussions and guidance. \newline


\section{Appendices}

\subsection{Fourier transform of the polaritonic Kagome Hamiltonian}

We now investigate the Hamiltonian of \cref{polar_fullhamilton} on the Kagome lattice geometry. The Hamiltonian acts on the $6N$-dimensional Hilbert space $\mathfrak H_N\otimes\mathfrak H_3\otimes \mathfrak H_2$. The $N$-dimensional space corresponds to the number of unit cells of the lattice, the 3-dimensional space denotes three internal levels of each unit cell, and the 2-dimensional space takes into account the polarization modes. It is straightforward to observe that the Zeeman term can be rewritten as

\begin{equation}
\hat H_{\textsf{kagome, (Zeem.)}}=\hat {\mathds 1}_N\otimes\hat {\mathds 1}_3\otimes\mathcal B\hat\sigma_z.
\end{equation}
The application of a magnetic field $\mathcal B$ allows to break TRS. For convenience, we split the full Hamiltonian into the following form
\begin{equation}
\hat H_{\textsf{kagome}} =\hat H_{\textsf{kagome,(Zeem.)}}+\hat H_{\textsf{kagome, (nn)}}+\hat H_{\textsf{kagome, (cp)}}.
\end{equation}
\justify

\begin{table}[h]
\centering
\begin{tabular}{l|l}
\hline
\hline
 Nomenclature & Meaning \\
\hline
$(m,n)$  &   unit cell in the Kagome lattice\\
$\sigma=\pm$ &   polarization mode \\
$\hat A/\hat A^{\dagger},\hat B/\hat B^{\dagger},\hat C/\hat C^{\dagger}$ & annihilation/creation operators for the levels in a unit cell \\
\hline
\hline
\end{tabular}
\caption{Labels for the operators of the Kagome lattice Hamiltonian.}
\end{table}
The problem can be mapped into the momentum space due to the periodicity of the lattice structure, which mathematically represents a Fourier transformation. For that purpose, we define Bloch states 
\begin{equation}
\ket{\textbf{k}}:=\frac{1}{\sqrt{N}}\sum_{(m,n)\in\mathbb Z^2}e^{i\textbf{k}\cdot\textbf{R}(m,n)}\ket{(m,n)}, \label{Bloch_state}
\end{equation}
where the sum runs over all unit cells and includes Kagome lattice vectors $\textbf{R}(m,n)$. The above expression can be used for other lattice geometries as well.

\paragraph{Nearest-neighbour-sum $\hat H_{\textsf{kagome}, (nn)}$ of the Hamiltonian.} 
The nn-hopping term of the Hamiltonian is ($\mathcal B=0$, $\delta t=0$)
\begin{equation}
\begin{split}
-\frac{1}{t}\hat H_{\textsf{kagome, (nn)}}=&\sum_{(m,n),\sigma}\hat A_{\sigma,(m,n)}^{\dagger}\hat B_{\sigma,(m,n)}+\hat A_{\sigma,(m,n)}^{\dagger}\hat C_{\sigma,(m,n)}+h.c. \\
 &+ \sum_{(m,n),\sigma}\hat B_{\sigma,(m,n)}^{\dagger}\hat C_{\sigma,(m,n)}+\hat B_{\sigma,(m,n)}^{\dagger}\hat A_{\sigma,(m+1,n)}+h.c.\\
 &+\sum_{(m,n),\sigma}\hat C_{\sigma,(m,n)}^{\dagger}\hat B_{\sigma,(m-1,n+1)}+\hat C_{\sigma,(m,n)}^{\dagger}\hat A_{\sigma,(m,n+1)}+h.c.\\
 &=\mathds 1_N\otimes 
\left(\begin{array}[pos]{ccc}
0 & 1 & 1\\
1 & 0 & 1\\
1 & 1 & 0
\end{array}\right)
\otimes \mathds 1_2 + \sum_{m,n}\ket{m,n}\bra{m+1,n}\otimes \ket{B}\bra{A}\otimes\mathds 1_2 +h.c. \\
& + \sum_{m,n}\ket{m,n}\bra{m-1,n+1}\otimes \ket{C}\bra{B}\otimes\mathds 1_2 + h.c.                  \\
& + \sum_{m,n}\ket{m,n}\bra{m,n+1}\otimes \ket{C}\bra{A}\otimes\mathds 1_2 + h.c.
\end{split}
\end{equation}
The Fourier transformed Hamiltonian is obtained from $\hat H(\textbf{k})_{\textsf{kagome, (nn)}}=\bra{\textbf{k}}\hat H_{\textsf{kagome, (nn)}}\ket{\textbf{k}}$ by inserting the Bloch states \eqref{Bloch_state}
\begin{equation}
\hat H(\textbf{k})_{\textsf{kagome, (nn)}}=-t
\left(\begin{array}[pos]{ccc}
0 & 1+e^{-i\textbf{k}\cdot \textbf{R}(1,0)} & 1+e^{-i\textbf{k}\cdot \textbf{R}(0,1)}\\
1+e^{i\textbf{k}\cdot \textbf{R}(1,0)} & 0 & 1+e^{-i\textbf{k}\cdot \textbf{R}(-1,1)}\\
1+e^{i\textbf{k}\cdot \textbf{R}(0,1)} & 1+e^{i\textbf{k}\cdot \textbf{R}(-1,1)} & 0
\end{array}\right)\otimes \mathds 1_2  .
\end{equation}

\paragraph{Cross-polarized term $\hat H_{\textsf{kagome, (cp)}}$ of the Hamiltonian.}
For the cross-polarized terms on the Kagome lattice we compute:
\begin{equation}
\begin{split}
&\sum_{m,n}\hat A^{\dagger}_{+,(m,n)}\hat B_{-,(m,n)}e^{i2\theta_{AB}} +\hat A^{\dagger}_{-,(m,n)}\hat B_{+,(m,n)}e^{-i2\theta_{AB}} +h.c.\\
&=\mathds 1_N\otimes\ket{A}\bra{B}\otimes\ket{+}\bra{-}e^{i2\theta_{AB}}+\mathds 1_N\otimes\ket{A}\bra{B}\otimes\ket{-}\bra{+}e^{-i2\theta_{AB}}+h.c.\\
&=\mathds 1_N\otimes(\ket{A}\bra{B}+\ket{B}\bra{A})\otimes
\underbrace{\left(\begin{array}[pos]{cc}
0 & e^{i2\theta_{AB}}\\
e^{-i2\theta_{AB}} & 0
\end{array}\right)}_{X_{AB}=X^{\dagger}_{AB}}
\end{split}
\end{equation}
The computation yields similar results with respect to pairs $(A,C)$ and $(B,C)$. The sum of these terms is 
\begin{equation}
\mathds 1_N\otimes 
\left(\begin{array}[pos]{ccc}
 0_2 & X_{AB} & X_{AC} \\
X_{AB} & 0_2 & X_{BC}\\
X_{AC} & X_{BC} & 0_2 
\end{array}\right). \label{cpterm1}
\end{equation}
We also need to evaluate the following sums 
\begin{gather}
\sum_{m,n}\ket{m,n}\bra{m+1,n}\otimes \left(\ket{B}\bra{A}\otimes \ket{+}\bra{-}e^{2i\theta_{AB}}+\ket{B}\bra{A}\otimes \ket{-}\bra{+}e^{-2i\theta_{AB}}\right) + h.c. \\
\sum_{m,n}\ket{m,n}\bra{m-1,n+1}\otimes \left(\ket{C}\bra{B}\otimes \ket{+}\bra{-}e^{2i\theta_{BC}}+\ket{C}\bra{B}\otimes \ket{-}\bra{+}e^{-2i\theta_{BC}}\right) + h.c.\\
\sum_{m,n}\ket{m,n}\bra{m,n+1}\otimes \left(\ket{C}\bra{A}\otimes \ket{+}\bra{-}e^{2i\theta_{AC}}+\ket{C}\bra{A}\otimes \ket{-}\bra{+}e^{-2i\theta_{BC}}\right) + h.c.
\end{gather}
The above expressions are now evaluated by squeezing them between Bloch states $\ket{\textbf{k}}$:
\begin{gather}
\left\{\ket{B}\bra{A}e^{i\textbf{k}\cdot \textbf{R}(1,0)}+e^{-i\textbf{k}\cdot \textbf{R}(1,0)}\ket{A}\bra{B}\right\}\otimes X_{AB}\\
\left\{\ket{C}\bra{B}e^{i\textbf{k}\cdot \textbf{R}(-1,1)}+e^{-i\textbf{k}\cdot \textbf{R}(-1,1)}\ket{B}\bra{C}\right\}\otimes X_{BC}\\
\left\{\ket{C}\bra{A}e^{i\textbf{k}\cdot \textbf{R}(0,1)}+e^{-i\textbf{k}\cdot \textbf{R}(0,1)}\ket{A}\bra{C}\right\}\otimes X_{AC}
\end{gather}
In total, performing the appropriate Fourier transformation on eq. \eqref{cpterm1}, and adding the above cross-polarized terms yields a $6\times 6$--matrix
\begin{equation}
\hat H(\textbf{k})_{\textsf{kagome, (cp)}}=-\delta t
\left(\begin{array}[pos]{ccc}
0_2 & (1+e^{-i\textbf{k}\cdot \textbf{R}(1,0)})X_{AB} & (1+e^{-i\textbf{k}\cdot \textbf{R}(0,1)})X_{AC} \\
(1+e^{i\textbf{k}\cdot \textbf{R}(1,0)})X_{AB} & 0_2 & (1+e^{-i\textbf{k}\cdot \textbf{R}(-1,1)})X_{BC}\\
(1+e^{i\textbf{k}\cdot \textbf{R}(0,1)})X_{AC} & (1+e^{i\textbf{k}\cdot \textbf{R}(-1,1)})X_{BC} & 0_2 
\end{array}\right)
\end{equation}

\paragraph{Complete Bloch Hamiltonian.} 
The Bloch Hamiltonian is the sum of the computed operators, $\hat H(\textbf{k})_{\textsf{kagome,(Zeem.)}}+\hat H(\textbf{k})_{\textsf{kagome, (nn)}}+\hat H(\textbf{k})_{\textsf{kagome, (cp)}}$, i.e.
\begin{equation}
\hat H(\textbf{k})=
\left(\begin{array}[pos]{ccc}
\mathcal B\hat\sigma_z & \mathcal M_{AB}(\textbf{k}) & \mathcal M_{AC}(\textbf{k}) \\
\mathcal M^{\dagger}_{AB}(\textbf{k}) & \mathcal B\hat\sigma_z & \mathcal M_{BC}(\textbf{k}) \\
\mathcal M^{\dagger}_{AC}(\textbf{k})  & \mathcal M^{\dagger}_{BC}(\textbf{k}) & \mathcal B\hat\sigma_z
\end{array}\right), \label{Bloch_kagome}
\end{equation}
with $2\times 2$-matrices defined by
\begin{equation}
\mathcal M_{\textbf{d}}(\textbf{k})=
-2e^{-i\textbf{k}\cdot \textbf{d}}\cos{(\textbf{k}\cdot \textbf{d})}
\left(\begin{array}[pos]{cc}
t & \delta t e^{2i\theta_{\textbf{d}}} \\
\delta t e^{-2i\theta_{\textbf{d}}}	& t
\end{array}\right).
\end{equation}

\newpage
\subsection{Numerical Algorithm: Computation of Chern numbers}
\justify

\begin{algorithm} 
(CHN-AL)
\begin{enumerate}
\item Define a grid $(i,j)$ on the surface $\mathcal S$; $i,j=1,\cdots, N$. \par \vspace{3mm}
\item For each node, solve the eigenvalue equation $\hat H(i,j)\ket{\Psi_n(i,j)}=E_n(i,j)\ket{\Psi_n(i,j)}$ with non-degenerate $E_n(i,j)$, and store eigenstates $\ket{\Psi_n(i,j)}$ in an array. \par \vspace{3mm}
\item For each square extract the 4 phases between the nearest neighbour eigenstates, e.g. $\gamma_{(i,j),(i+1,j)}=-\arg(\braket{\Psi_n(i,j)|\Psi_n(i+1,j)})$. Subsequently, build the sum $\Gamma_{ij}$ as in \cref{phase_sum} and compute $\Omega^{(n)}_{ij}=-\arg(\exp(-i\Gamma_{ij}))$.\par \vspace{3mm}
\item Sum all fluxes $\Omega^{(n)}_{ij}$ through the $(N-1)^2$ squares to obtain the Chern number $\mathcal C^{(n)}=\frac{1}{2\pi}\sum_{ij}\Omega^{(n)}_{ij}$ which corresponds to the nth level.\par \vspace{3mm}
\end{enumerate} \label{CHN_AL}
\end{algorithm}
\justify 

As part of this work, the algorithm has been efficiently implemented in the Python programming language using its NumPy library for numerical operations on high-dimensional arrays. As anticipated by the results of Fukui et al. \cite{Fukui}, we can confirm the fast convergence of this method, even for a coarse discretization of the underlying space.

\begin{figure}[h]
\centering
\begin{tikzpicture}[roundnode/.style={circle, draw=blue!90, fill=gray!7, thin, scale=0.5}] 
\draw[step=1cm,blue!30, very thick] (-1.9,-1.9) grid (4.9,4.9);
\node (1) at (0,0) [roundnode] {$\Psi_1$};
\node (2) at (0,1) [roundnode] {$\Psi_2$};
\node (3) at (1,1) [roundnode] {$\Psi_3$};
\node (4) at (1,2) [roundnode] {$\Psi_4$};
\node (5) at (2,2) [roundnode] {$\Psi_5$};
\node (6) at (2,3) [roundnode] {$\Psi_6$};
\node (7) at (3,3) [roundnode] {$\Psi_7$};
\node (8) at (4,3) [roundnode] {$\Psi_8$};
\node (9) at (4,2) [roundnode] {$\Psi_9$};
\node (10) at (4,1) [roundnode] {$\Psi_{10}$};
\node (11) at (3,1) [roundnode] {$\Psi_{11}$};
\node (12) at (2,1) [roundnode] {$\Psi_{12}$};
\node (13) at (2,0) [roundnode] {$\Psi_{13}$};
\node (14) at (1,0) [roundnode] {$\Psi_{14}$};
\draw[->, black!,  thick] (1) to (2);
\draw[->, black!,  thick] (2) to (3);
\draw[->, black!,  thick] (3) to (4);
\draw[->, black!,  thick] (4) to (5);
\draw[->, black!,  thick] (5) to (6);
\draw[->, black!,  thick] (6) to (7);
\draw[->, black!,  thick] (7) to (8);
\draw[->, black!,  thick] (8) to (9);
\draw[->, black!,  thick] (9) to (10);
\draw[->, black!,  thick] (10) to (11);
\draw[->, black!,  thick] (11) to (12);
\draw[->, black!,  thick] (12) to (13);
\draw[->, black!,  thick] (13) to (14);
\draw[->, black!,  thick] (14) to (1);
\end{tikzpicture}
\caption{Construction of a Berry phase. A sample loop $\mathcal C$ which connects states on a grid configuration.}
\label{fig:disc_berryph}
\end{figure}
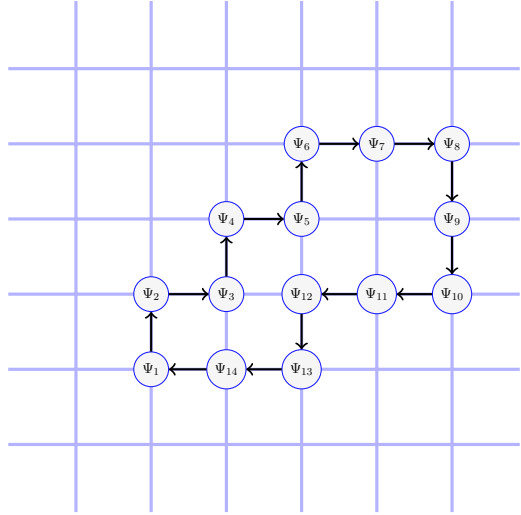
Note the phase sum taken across the boundary of each plaquette is given by (see \cref{fig:disc_berryph})
\begin{equation}
\Gamma_{ij}:=\gamma_{(i,j),(i+1,j)}+\gamma_{(i+1,j),(i+1,j+1)}+\gamma_{(i+1,j+1),(i,j+1)}+\gamma_{(i,j+1),(i,j)}. \label{phase_sum}
\end{equation}

\newpage
\subsection{Symmetries of Hamiltonians}

The symmetries discussed here underlie the classification of topological insulators. 
A symmetry of a Hamiltonian $\hat H$ is given by a map $\hat H\to \hat S\hat H\hat S^{\dagger}$ where $\hat S$ can be either a unitary or anti-unitary operator due to Wigner's theorem (see \cite{Molnar} for a proof). Anti-unitary operators can be always written as a product $\hat U K$ where $\hat U$ is unitary and $K$ denotes complex conjugation operation. 

\subsubsection{Time Reversal Symmetry (TR)}

TR is given by the transformation $t\to -t$. If the \textit{real space Hamiltonian} $\hat H$ is invariant under TR, then, using the Schrödinger equation one sees immediately that the time-reversal operator must be anti-unitary
\begin{equation}
 \hat {\mathcal T}=\hat U_\textsf{T} K,\quad \hat {\mathcal T}\hat {\mathcal T}^{\dagger}=\hat{\mathds 1},\quad  \hat{\mathcal T}\hat H{\mathcal T}^{\dagger}=\hat H. \label{TRS_P1}
\end{equation}
We derive some properties. Consecutive operation of $\hat H$ and $\hat{\mathcal T}$ on an arbitrary state $\ket{\psi}$ yields 
\begin{equation}
\hat{\mathcal T}\hat H\ket{\psi}=\hat U_\textsf{T} K\hat H\ket{\psi}=\hat U_\textsf{T}\hat H^{*}\ket{\psi}^*=\hat U_\textsf{T}\hat H^{*}\hat U^{\dagger}_\textsf{T}\hat U_\textsf{T}K\ket{\psi}=\hat U_\textsf{T}\hat H^{*}\hat U^{\dagger}_\textsf{T}\hat{\mathcal T}\ket{\psi}
\end{equation}
\begin{equation}
\Rightarrow \hat{\mathcal T}\hat H{\mathcal T}^{\dagger}=\hat U_\textsf{T}\hat H^{*}\hat U^{\dagger}_\textsf{T}=\hat H.
\end{equation}
For the momentum-space Hamiltonian $\hat H(\textbf{k})$ the transformation $t\to -t$ leads to $\textbf{k}\to -\textbf{k}$:
\begin{equation}
\hat{\mathcal T}\hat H(\textbf{k}){\mathcal T}^{\dagger}=\hat U_\textsf{T}\hat H^{*}(\textbf{k})\hat U^{\dagger}_\textsf{T}=\hat H(-\textbf{k}).
\end{equation}
Applying the $\hat{\mathcal T}$-operator twice to a state must return a physically equivalent state - i.e. $\hat{\mathcal T}^2=e^{i\varphi}$. So, $\hat U_\textsf{T}K\hat U_\textsf{T}K=\hat U_\textsf{T}\hat U^{*}_\textsf{T}=e^{i\varphi}$. Thus,
\begin{equation}
\hat U_\textsf{T}\underbrace{\hat U^{*}_\textsf{T}\hat U_\textsf{T}}_{e^{-i\varphi}}\hat U^{*}_\textsf{T}=e^{i2\varphi}
\Rightarrow e^{-i\varphi}\hat U_\textsf{T}\hat U^{*}_\textsf{T}=\hat{\mathds 1}=e^{i2\varphi}.
\end{equation}
This implies $e^{i\varphi}=\pm 1$ as the only two possibilities or, equivalently $\hat{\mathcal T}^2=\pm \hat{\mathds 1}$.

\begin{lemma} 
(Kramer's degeneracy)
Let $\hat H\ket{\psi}=E\ket{\psi}$ be the eigenvalue equation and $[\hat H,\hat{\mathcal T}]=0$ such that $\hat{\mathcal T}^2=-\mathds 1$ (fermionic condition) holds. Then the states $\ket{\psi}$ and $\hat{\mathcal T}\ket{\psi}$ are orthogonal,
\begin{equation}
\braket{\psi|\hat{\mathcal T}\psi}=0. \label{Kramer}
\end{equation}
Hence, each Hilbert space $\mathfrak H_E$ is at least double degenerate, $\dim \mathfrak H_E\geq 2$.
\end{lemma}
\begin{proof}
$\hat{\mathcal T}\ket{\psi}$ is an eigenstate to $E$ since $[\hat H,\hat{\mathcal T}]=0$. We write 
\begin{equation*}
\braket{\psi|\hat{\mathcal T}\psi}=\braket{\psi|\hat{\mathcal T}^{\dagger}\underbrace{\hat{\mathcal T}\hat{\mathcal T}}_{-\mathds 1}\psi}
=-\braket{\psi|\hat{\mathcal T}^{\dagger}\psi}=-\braket{\hat{\mathcal T}\psi|\psi}^{*}=-\braket{\psi|\hat{\mathcal T}\psi},
\end{equation*}
where use has been made of the fact that $\hat{\mathcal T}$ is anti-unitary. Hence, $\braket{\psi|\hat{\mathcal T}\psi}=0$.
\end{proof}

\subsubsection{Particle-Hole Symmetry (PH)} 

PH symmetry stems from the existence of particles (occupied states/sites) and holes (unoccupied states/sites). It is represented by an anti-unitary operator, however, with the difference that it anticommutes with the Hamiltonian due to the fact that unoccupied states carry the opposite energy of the occupied ones.
\begin{equation}
\hat {\mathcal P} \hat H\hat {\mathcal P}^{\dagger}=-\hat H
\end{equation}
In the same way as for TR symmetry we get, by setting $\hat{\mathcal P}=\hat U_\textsf{P}K$,
\begin{equation}
\hat {\mathcal P} \hat H\hat {\mathcal P}^{\dagger}=\hat U_\textsf{P}\hat H^{*}\hat U^{\dagger}_\textsf{P}=-\hat H.
\end{equation}
The effect on the Fourier transformed Hamiltonian is 
\begin{equation}
\hat {\mathcal P} \hat H(\textbf{k})\hat {\mathcal P}^{\dagger}=\hat U_\textsf{P}\hat H^{*}(\textbf{k})\hat U^{\dagger}_\textsf{P}=-\hat H(-\textbf{k}).
\end{equation}
The crucial point is an effect on the energy spectrum or band structure of $\hat H(\textbf{k})$. Assume you have a (positive) band $E^{(+)}(\textbf{k})>0$, then there must exist a (negative) band $E^{(-)}(\textbf{k})$ such that $E^{(-)}(-\textbf{k})=-E^{(+)}(\textbf{k})$ holds (reflection symmetry with respect to the zero point). This can be used as a check for PH symmetry once the full band structure of the system has been computed or measured.

\subsubsection{Sublattice Symmetry (SL)}

SL symmetry is also known, depending on the context, as chiral symmetry. The corresponding, now unitary operator $\hat{\mathcal S}$, anticommutes with $\hat H$ and the symmetry condition is
\begin{equation}
\hat {\mathcal S} \hat H\hat {\mathcal S}^{\dagger}=-\hat H, \label{sub_latt_sym}
\end{equation}
as e.g. in the SSH lattice. Forming the product $\hat {\mathcal P}\cdot\hat {\mathcal T}$ is one way of constructing such an $\hat{\mathcal S}$ operator. This implies that a system with both TR and PH symmetry has also SL symmetry. At the momentum-space level we have 
\begin{equation}
\hat {\mathcal S} \hat H(\textbf{k})\hat {\mathcal S}^{\dagger}=-\hat H(\textbf{k}).
\end{equation}
Let $\hat {\mathcal S}=\hat {\mathcal P}\cdot\hat {\mathcal T}=\hat U_\textsf{P}\hat U^{*}_\textsf{T}$, then $\hat {\mathcal S}\hat H(\textbf{k})\hat {\mathcal S}^{\dagger}=\hat {\mathcal P}\hat {\mathcal T}\hat {H}(\textbf{k})\hat {\mathcal T}^{\dagger}\hat {\mathcal P}^{\dagger}=\hat {\mathcal P}\hat {H}(-\textbf{k})\hat {\mathcal P}^{\dagger}=-\hat H(\textbf{k})$, which demonstrates that SL symmetry follows from TR and PH symmetry. However, it should be noted that the reverse statement is not true in general - i.e. there can exist systems with SL symmetry but no TR and no PH symmetry.
The effect of SL symmetry on $\hat H(\textbf{k})$ is a fully symmetric band structure about $\mathcal {BZ}$. If $E^{(+)}(\textbf{k})>0$ is a positive band, then there exists a negative band $E^{(-)}(\textbf{k})$ such that $E^{(-)}(\textbf{k})=-E^{(+)}(\textbf{k}), \forall \textbf{k} \in \mathcal{BZ}$.

\subsubsection{Altland-Zirnbauer classification}

\begin{table}[H]
\centering
\begin{tabular}{r|r|r|l|r}
\hline
\hline
 \textbf{TRS} & \textbf{PHS} & \textbf{SLS} & \textbf{Cartan label} & \textbf{Coset space of Hamiltonian} \\
\hline
\hline
 $0 $   &  $0$  &   $0$ & A      &  $U(n+m)/(U(n)\times U(m))=\mathbb G_{m,n+m}(\mathbb C)$ \\

 $1$    & $0$   &   $0$ & AI     &  $O(n+m)/(O(n)\times O(m))=\mathbb G_{m,n+m}(\mathbb R)$\\

$-1$    & $0$   &  $0$ & AII      &  $Sp(n+m)/(Sp(n)\times Sp(m))$ \\
\hline
$0$     &  $0$    &  $1$  &  AIII  & $(U(n)\times U(n))/U(n)$   \\

$1 $  &    $1$  &   $1$ &     BDI   &    $(O(n)\times O(n))/O(n)$    \\
$-1$   &  $-1$  &    $1$ &    CII   &   $(Sp(n)\times Sp(n))/Sp(n)$  \\
\hline
 $0$   &    $1$  &    $0$  &     D  &     $O(2n)/U(n)$                \\
 $0 $  &   $-1$  &    $0$  &     C  &     $Sp(2n)/U(n)$                \\
 $-1$  &    $1$  &    $1$  &     DIII  &  $U(2n)/Sp(2n)$                \\
 $1$ &     $-1$  &   $1$   &     CI  &    $U(n)/O(n)$                    \\
\hline
\hline
\end{tabular}
\caption{"10-fold way". Altland-Zirnbauer table \cite{Altland, Heinzner, Kitaev} for the ten symmetry classes of Hamiltonians according to time-reversal (TR), particle-hole (PH) and sublattice (SL) symmetry. The numbers $0,\pm 1$ denote absence, presence of the symmetry, respectively. Moreover, $\pm 1$ refers to the properties $\hat{\mathcal T}^2=\pm \mathds 1$ and $\hat{\mathcal P}^2=\pm \mathds 1$. Note that the cosets are homogeneous spaces, represented as quotients $G/H$ of two Lie groups $G$ and $H$.}
\label{fig:Altl_Zirn}
\end{table}

\clearpage
\bibliography{articlebib}

\end{document}